\documentclass[12pt, epsfig]{article}     

\usepackage{amssymb}
\usepackage{latexsym}
\usepackage{amscd}

\usepackage{amsmath}
\usepackage{amsfonts}
\usepackage{eucal}
\usepackage{epsfig}

\newcommand{\beal}{\begin{subequations} \begin{align}}
\newcommand{\eeal}{\end{align} \end{subequations}}
\newcommand{\be}{\begin{equation}}
\newcommand{\ee}{\end{equation}}
\newcommand{\bsp}{\begin{split}}
\newcommand{\esp}{\end{split}}
\newcommand{\bea}{\begin{eqnarray}}
\newcommand{\eea}{\end{eqnarray}}
\newcommand{\ba}{\begin{array}}
\newcommand{\ea}{\end{array}}

\newcommand{\Tr}{{\mbox{Tr}}}

\makeatletter
\@addtoreset{equation}{section}

\makeatother

\def\ie{{\it i.e.} }
\def\cf{{\it cf.} }
\def\hmu{\hat{\mu}}
\def\hnu{\hat{\nu}}

\def\bg{background }

\newcommand{\gd}{\delta}

\newcommand{\half}{\frac12}

\begin{document}

\thispagestyle{empty}

\begin{flushright}
\vspace*{1mm}
\hfill{hep-th/0406053}\\
\hfill{SU-ITP-04/23}
\end{flushright}

\vskip 1cm

\begin{center}

{\Huge \bf S}{\huge\bf q}{\LARGE\bf u}{\Large\bf a}{\Large\bf s}{\LARGE\bf h}{\huge\bf 
e}{\Huge\bf d} {\Huge\bf Giants:}
\vskip .5cm
{\Large\bf Bound States of Giant Gravitons}

\vskip 1.0cm

{\large \sc Sergey Prokushkin and M.M. Sheikh-Jabbari}

\vskip 1cm

\medskip
{\it Department of Physics, Stanford University,}
{\it Stanford, CA 94305, USA}\\
{\tt prok, jabbari@itp.stanford.edu}

\end{center}

\vskip 2cm

\begin{abstract}

We consider giant gravitons in the maximally supersymmetric type IIB plane-wave, in the
presence of a constant NSNS $B$-field background. We show that in response to the background
$B$-field the giant graviton would take the shape of a deformed three-sphere, the size and
shape of which depend on the $B$-field, and that the giant becomes classically unstable once 
the $B$-field is larger than a critical value $B_{cr}$.  In particular, for the $B$-field which
is (anti-)self-dual under the $SO(4)$ isometry of the original giant $S^3$, the closed string
metric is that of a round $S^3$, while the open string metric is a squashed three-sphere.  
The squashed giant can be interpreted as a bound state of a spherical three-brane and 
circular D-strings.  We work out the spectrum of geometric fluctuations of the squashed giant 
and study its stability . We also comment on the gauge theory which 
lives on the
brane (which is generically a noncommutative theory) and a possible dual gauge theory
description of the deformed giant.

\end{abstract}

\newpage

\section{Introduction}

Usual D$p$-branes are $p+1$ dimensional objects which carry one unit of RR $p+1$ form charge
and have tension $(l_s^{p+1} g_s)^{-1}$. One might try to construct $p$-branes whose 
worldvolume is (topologically) $R\times S^p$. Evidently such branes, unlike flat D$p$-branes, 
cannot carry a net $p+1$ form RR charge. Moreover, in the absence of any other force acting on 
such branes, they would immediately collapse under their tension. Although a  spherical brane 
cannot carry a net RR charge, it has  (electric) dipole moment of such charges. In this 
respect such branes behave similarly to the usual fundamental strings and in 
particular supergravity states, hence these spherical branes were called {\it giant 
gravitons} \cite{McGreevy:2000cw}. One can use this dipole moment to exert a force on the 
brane which 
cancels off the tension force and stabilize the brane at a finite size. This can be done if the 
giant graviton is moving in a background with a non-zero (magnetic) flux of the corresponding 
$p+1$ form, and the size of the giant $R_0$ is related to  the (angular) momentum $J$
as $R_0^{p-1}\propto J$ \cite{McGreevy:2000cw}. So, to stabilize the giant we need two basic 
ingredients: background $p+1$ form flux and a moving brane. The simplest and most famous 
examples of 
such backgrounds are $AdS_p\times S^q$ geometries with $(p,q)= (5,5),\ (4, 7)$ or $(7,4)$.
 Various aspects of the giants in these backgrounds have been studied in 
\cite{McGreevy:2000cw, Grisaru:2000zn, Hashimoto:2000zp, Das:2000st, Balasubramanian:2001nh, 
Balasubramanian:2002sa}.

Besides the $AdS$ backgrounds, recently the plane-wave backgrounds have also been under 
intense 
study, for a review see \cite{juan, review}. Plane-waves, as solutions of supergravity 
generically have a non-vanishing form flux and  these fluxes can be used to 
stabilize spherical branes. The plane-wave background that we would focus on here is the 
maximally supersymmetric type IIB background (we follow the notations and conventions of 
\cite{review}): 
\begin{subequations}\label{background}
\begin{align}
  ds^2  =  -2 dX^+ dX^- & -\mu^2(X^i X^i + X^a X^a) {(dX^+)}^2 + dX^i dX^i+ dX^a dX^a \, ,\\
  F_{+ijkl} &= \frac{4}{g_s} \mu\ \epsilon_{ijkl} \, ,\ \ \ \ \ \ 
  F_{+abcd}=\frac{4}{g_s} \mu \ \epsilon_{abcd} \ , 
\end{align}
\end{subequations}
where $i,a=1,2,3,4$.
This background has a globally defined light-like Killing vector $\partial/\partial X^-$.  
As discussed in \cite{Hh} this background admits a stable three sphere giant graviton 
solution 
with the worldvolume along $X^+$ direction (the light-cone time) and three sphere is embedded 
in either the $X^i$ or $X^a$ directions. In particular note that $X^-$ is transverse to the 
giant.

String theory $\sigma$-model on the plane-wave background is shown to be solvable in the 
light-cone gauge \cite{review, MT}. In the light-cone gauge $X^+=\tau$ (where $\tau$ is the 
worldsheet time) and $X^-$ is a non-dynamical variable, completely determined through the 
transverse $X^i$ and $X^a$ directions, in particular \cite{review}
\be\label{level-matching}
\partial_\sigma X^-= \partial_\sigma X^i \partial_\tau X^i+\partial_\sigma X^a 
\partial_\tau X^a\ ,
\ee
where  $\sigma, \tau$ parametrize the worldsheet.
One might then wonder whether giant gravitons, similarly to  ordinary D-branes, have a
perturbative description in terms of open strings ending on them with Dirichlet boundary 
conditions along the directions transverse to the brane. Noting \eqref{level-matching}
one would readily see that, independently of the boundary conditions on $X^i$ or $X^a$ 
directions, $X^-$ would satisfy Neumann boundary condition. This implies that $X^-$ should be
along the brane, while in our giant graviton $X^-$ is transverse to the brane. Therefore, 
giant 
graviton does not have a simple open string description and lots of the properties of the 
usual D-branes, e.g. the fact that the low energy effective field theory on a single three 
sphere giant is a $U(1)$ supersymmetric gauge theory on $R\times S^3$ and that for $K$
coincident giants the gauge symmetry enhances to $U(K)$, if correct, are harder to realize or 
argue for.
Although there is no simple open string picture for giants,
studying the theory residing on the giant graviton,  it was shown that
there are (spike-type) BPS solutions on the giant three sphere which have the same physical
behaviour one expects form the open strings ending on a spherical D-brane \cite{Hh}; \ie
giant gravitons are spherical D-branes.

Here we study behaviour of a giant in the plane-wave \eqref{background} when a constant NSNS
$B$-field is turned on in the background. Turning on a constant NSNS background field
would not change the geometry, as in the supergravity equations of motion only $H=dB$ 
appears which is vanishing in our case. Note that, existence of the selfdual five form flux 
in the background does not change this result. However,   presence of the 
background five form flux affects dynamics of NSNS and RR two form fields so that different 
polarizations of the two form fields have different light-cone masses \cite{ review, MT}. One 
of the main motivations for studying this 
problem is that it may help with solving the long-standing problem of quantizing a $p$-brane
($p>1$). Giant gravitons are particularly nice laboratories for attacking this problem mainly 
because, unlike the usual D$p$-branes, their worldvolume is naturally compact and they have a 
finite volume and also the spectrum of fluctuations of the giant is discrete and (in the 
free field theory limit) is given by equally spaced integers \cite{Hh}. 

For a flat D$p$-brane, where a simple perturbative open string description is available, 
turning on the $B$-field along the brane simply amounts to replacing the Neumann boundary 
conditions with a mixed (Neumann and Dirichlet) boundary condition along the directions of 
the 
$B$-field \cite{Bound}. As a result a D$p$-brane in the $B$-field background behaves as  
a bound state of  D$p$ and lower dimensional branes \cite{Bound} and the low energy 
effective field theory on 
the brane is now a $p+1$ dimensional noncommutative gauge theory, e.g. see \cite{SW}.
For the giant graviton case, however, we do not have the simple open string description and
hence our analysis is more limited to the Born-Infeld action and supersymmetry algebra.

As mentioned earlier, spherical shape of the giant is a result of the balance between 
tension and forces coming from the form fluxes. As we will show in section 2, in the presence 
of the background $B$-field this balance is lost and hence the giant needs to 
reshape itself to adjust to the presence of $B$-field so that the  shape of the giant is now a 
deformed three-sphere, the ``squashed giant''. This reshaping, which of course has no 
counterpart in the flat brane case, among the other things, would provide us with a chance 
of quantizing a submanifold of $S^3$ worldvolume, explicitly an $S^2\in S^3$. This point 
will be addressed in some detail in section 2.5, where we show that it leads to the
novel feature of quantization  of the $B$-field. As we will show, if the $B$-field is larger 
than some critical value $B_{cr}$ the giant cannot adjust itself to the $B$-field and
becomes unstable. In other words, the squashed giant state only exists for $B<B_{cr}$.
Once we calculated the shape of the deformed giant we look for a physical   interpretation 
for the deformed or ``squashed'' giant. We will argue in section 2.6, that the squashed giant 
is indeed a bound state of a giant spherical D3-brane and circular D-strings wrapping on the 
$S^3$. 

After establishing what the squashed giant is in section 2, in sections 3 and 4 we 
address the question of its stability. In section 3, we 
analyze   small fluctuations of the giant and study  corrections to the spectrum due to 
the deformation of the shape and presence of the $B$-field. In section 4, we show that 
squashed giant is a 1/4 BPS object. We use the supersymmetry analysis to argue that squashed 
giant is classically stable, however, quantum mechanically, through instanton effects, it 
would decay into the zero size branes and usual supergravity modes. We close the paper 
by a summary of our results, a proposal for a possible description of squashed giants in 
the dual ${\cal N}=4$ gauge theory as well as a list of interesting questions which we did not 
address in this work.

\section{Giant gravitons in a constant $B$-field background}

In this section we first present the light-cone  Hamiltonian of a 3-brane in the plane-wave 
\eqref{background} with a constant {\it magnetic}
background $B$-field, i.e. the $B$-field has no legs along the $X^+$, and $X^-$ directions.
Without loss of generality such $B$-field can be chosen to be along $X^i$ directions. In 
other words we choose the only non-zero components of the $B$-field to be  
$B_{ij}$ components.
In section \ref{Reshape}, we show that the round 
three sphere solution is no longer minimizing the potential, and in response to 
the background $B$-field the giant graviton takes a new shape. 
Giant gravitons in the non-constant background $B$-field and 
non-spherical giants, though in a different context, have been considered 
in \cite{Lunin, Mikhailov}. 
In section~\ref{r-lambda-potential}, 
we analyze the potential in some detail and compute the 
new shape of giant graviton which minimizes the potential. In  section 
\ref{ASD}, we consider a particularly interesting example of a $B$-field
background which preserves $SU(2)\times U(1)$ subgroup of $SO(4)$ isometry of the three-sphere
(or the background plane-wave) and finally in section 2.6, we discuss that the ``squashed'' 
giant is indeed a bound state of a three sphere giant and circular D-strings.

\subsection{Light-cone Hamiltonian of giants in the plane-wave}\label{LCH}

To study a 3-brane in the plane-wave background we start with the Born-Infeld action:
\be \label{DBI-action}
  S=  \frac{1}{l_s^4g_s}\int d\tau d^3\sigma \:  
  \sqrt{-det \left( g_{\hmu\hnu} + {\cal F}_{\hmu\hnu} \right)} +\int C_4 
+\int C_2\wedge{\cal F}+\int\frac{1}{2}\chi {\cal F}\wedge{\cal F} ,
\ee
where $\hmu,\hnu=0,1,2,3$ indices correspond to the  worldvolume coordinates $\tau, 
\sigma^r$, $r=1,2,3$. $g_{\hmu\hnu}$ is the induced metric on the brane:
\[
g_{\hmu\hnu}=G_{\mu\nu}\partial_{\hmu} X^{\mu}\partial_{\hnu} X^{\nu}
\]
where $G_{\mu\nu}$ is the background plane-wave metric and 
$X^{\mu}=(X^+,X^-, X^i, X^a)$ are the embedding coordinates of the three-brane.
$C_4, C_2$ and $\chi$ are the background RR form fluxes. In our case we only have a non-zero 
$C_4$, whose field strength is the background five form flux of (\ref{background}b), 
explicitly
\be\label{fourform}
  C_{+ijk} = -\frac{\mu}{g_s} \epsilon_{ijkl} X^l \, , \ \ \ \ 
  C_{+a b c} = -\frac{\mu}{g_s} \epsilon_{a b c d} X^{d} \, .
\ee
${\cal F}_{\hmu\hnu}=b_{\hmu\hnu}+ F_{\hmu\hnu}$ is the invariant $U(1)$ field of the brane, 
$F=dA$, $A$ is the $U(1)$ gauge field on the brane and  $b$ is the pullback of the 
background  NSNS two form field
\[
          b_{\hmu\hnu}=B_{\mu\nu}\partial_{\hmu} X^{\mu}\partial_{\hnu} X^{\nu}.
\]
  
In the plane-wave background, due to the existence of a globally defined null Killing vector 
field \cite{review}, it is particularly useful to fix the light-cone gauge by taking
\footnote{Here we do not repeat  details of the light-cone gauge fixing, for 
which the reader is referred to \cite{Hh}.}
\[
          X^+=\tau, \ \ \ \qquad g_{\tau\sigma_r}=0\ .
\]   
Following the analysis of \cite{Hh} we obtain the light-cone Hamiltonian { density} in the 
presence of a non-zero NSNS $B$-field: 
\be\label{Hlc}
        {H}_{l.c.} = \frac{1}{2 p^+}P^I P^I+ V(X^i,X^a) \, ,
\ee
with
\bea\label{potential}
         V(X^i,X^a)&=&\frac{\mu^2 p^+}{2}(X_i^2+X_a^2)+\frac{1}{2 p^+g_s^2} \; \mbox{det} \: 
              (g_{rs} + b_{rs}) \cr  &-&\frac{\mu}{6g_s} \Bigl(
              \epsilon^{i j k l} X^i\{ X^j, X^k, X^l \}+ \epsilon^{a b c d} X^a\{ X^b, X^c, 
X^d \} \Big) \ , 
\eea
where
$g_{rs}$ and  $b_{rs}$ are the spatial parts of pullbacks of the background metric and 
$B$-field onto the spatial part of the brane worldvolume. For our choice of the $B$-field
\be
        g_{rs}=\partial_r X^i\partial_s X^i+\partial_r X^a\partial_s X^a \ ,\quad 
        b_{rs}=\partial_r X^{i}\partial_s X^{j} \;  B_{ij} \,, \quad r,s = 1,2,3 \,,
\ee
and 
\be\label{Nambubracket}
         \{ F , G , K \} = \epsilon^{p r s} \partial_p F \partial_r G \partial_s K \, ,
\ee
is the Nambu bracket, where the antisymmetrization is with respect to the worldvolume 
coordinates $\sigma^r$. Using the definition of determinant, the Nambu bracket  \eqref{Nambubracket} 
and the fact that $B_{ij}$ is antisymmetric under exchange of $i$ and $j$ one can show that 
\bea\label{det}
     \mbox{det} (g_{rs} + b_{rs}) &=&  \frac{1}{6} \{ X^i , X^j , X^k \} \{ X^i , X^j , X^k \} 
+\frac{1}{6} \{ X^a , X^b , X^c \} \{ X^a , X^b , X^c \} \cr &+& 
\frac{1}{2} \{ X^i , X^j , X^a \} \{ X^i , X^j , X^a \}  
+\frac{1}{2} \{ X^i , X^a , X^b \} \{ X^i , X^a , X^b \}\ \\ 
&+& {\half}  \left(\{ X^i , X^k , X^m \} \{ X^j , X^l , X^m \}+\{ X^i , X^k , 
X^a \} \{ X^j , X^l , X^a \}\right)\; B_{ij} B_{kl}\nonumber .  
\eea
The first two line of \eqref{det} is nothing but $\det g_{rs}$.  
As we see the whole Hamiltonian can nicely be written in terms of the Nambu brackets.

In the case with $B=0$, it was shown that \cite{Hh}  there are three minimum energy, half 
BPS solutions: one is point-like $X^i=X^a=0$, and the other two  are  spherical three-branes, 
grown along the $X^i$ directions and sitting at $X^a=0$ or   
grown along the $X^a$ directions and sitting at $X^i=0$.  The latter two solutions are 
related 
by the $Z_2$ symmetry which exchanges the $X^i$ and $X^a$ directions \cite{Hh}. Here we focus 
on 
the $X^a=0$ solutions. In the absence of the $B$-field and setting $X^a=0$, the above 
potential is minimized at
\be\label{b=0vacua}
\epsilon_{ijkl}\{ X^j, X^k, X^l\}=6 g_s\mu p^+ X^i\ .
\ee
The finite size solution of \eqref{b=0vacua} can be expressed through
\be\label{spheric}
         X^i = R_0 x^i \,,  \qquad R_0^2 =  \mu p^+g_s ,
\ee 
where $x^i$'s, which satisfy
\be\label{units3}
         x^ix^i = 1 \,, \quad  \{x^j,x^k,x^l\}=\epsilon^{ijkl}x^i\,,  
\ee
are the embedding coordinates of a unit three-sphere in $\mathbb{R}^4$. 

\subsection{\!Reshaping:\! Response of the giant graviton to the $B$-field}\label{Reshape}

Now we consider configurations with a non-zero, but constant (i.e. $dB=0$)  background 
$B$-field. 
Although a constant background  $B$-field does not affect the (perturbative) dynamics of 
the closed strings, D-branes (giant gravitons) would feel the presence of the $B$-field which 
has both of its legs along the brane \cite{SW}. 
Using \eqref{det}, it is straightforward to see that in the presence of the $B_{ij}$ field, 
the three sphere giant grown in $X^a$ directions (sitting at $X^i=0$) is still a 
zero energy solution of the light-cone Hamiltonian \eqref{Hlc}. 
Therefore, with the choice of the $B_{ij}$ field, 
we focus on the giant which is grown in the $X^i$ directions. 
Setting $X^a=0$,  the potential (\ref{potential}) can be rewritten as
\bea\label{V}
   V(X^i, X^a=0)&=& \frac{1}{2p^+}\left(\mu p^+ X^i-\frac{1}{6g_s}                              
  \epsilon_{ijkl}\{ X^j, X^k, X^l\}\right)^2 \cr                         
&+& \frac{1}{4p^+ g_s^2}\,\{ X^i , X^k , X^m \} \{ X^j , X^l , X^m \}\; B_{ij} B_{kl}  \,.
\eea

To find how a spherical three-brane responds to this $B$-field, 
let us consider fluctuations of the embedding coordinates around the spherical solution 
(\ref{spheric}),
\be\label{round}
            X^i = R_0\, x^i + Y^i \,.   
\ee
Plugging this ansatz into  (\ref{V})  one gets the potential in the  form of the expansion
\footnote{Note that the total potential energy of the brane is an integral of $V$ over the 
spatial part of the brane worldvolume $M_3$: 
\[
E_p = \int_{M_3}\, d^3 \sigma \;V(\sigma) = \int_{\mathbb{R}^4} d^4 x \; \gd(x^ix^i - 1)\; 
V(x)  \,.
\]
We will use this fact and by-parts integration to obtain different expansion 
terms in (\ref{Yexpan}) and \eqref{V}.} 
\be\label{Yexpan}
            V = V^{(0)} + V^{(1)}_i \, Y^i + V^{(2)}_{ij}\, Y^i Y^j + V^{(3)}_{ijk}\, Y^i 
Y^j Y^k  + O[(Y^i)^4] \,.
\ee

The $V^{(0)}$ term, which is the zero point energy, is zero in the $B=0$ case. This can be 
easily 
seen from the first line of \eqref{V}, and is a result of the fact that the round 
three-sphere giant 
is a half SUSY state with zero light-cone energy \cite{Hh}. In the $B\neq 0$ case,  
\[
           V^{(0)}=\frac{1}{4 p^+ g_s^2}R_0^6\left(\frac{1}{2} B^2+ 2T_{ij}(B)x^ix^j\right)
\]
where   
\be\label{T}
           B^2 = B_{ik}B_{ik}\ , \qquad   T_{ij}(B) = B_{ik}B_{kj} + \frac{1}{4}\delta_{ij} 
B^2\ .   
\ee
Note that $T_{ij}$ is a symmetric traceless tensor ({\it i.e.} it lies in ${\bf 9}$ of $SO(4)$).
The potential \eqref{V} is a density, and to obtain contribution of $V^{(0)}$ to the total 
energy we need to integrate the potential over the unit three-sphere volume. Noting 
that\footnote{We have defined the measure of the integral so that $\int_{s^3} d\Omega_3=1$.}
\be\label{xixj-integral}
           \int_{s^3} d\Omega_3\  x^i x^j=\frac{1}{4}\delta_{ij}
\ee 
and choosing a constant $B$-field, {\it i.e.} $\{x^i, x^j, B_{kl}\}=0$, we find
\be\label{V(0)}
\begin{split}
           V^{(0)}&=\frac{1}{8 p^+ g_s^2} R_0^6 B^2=\mu\cdot \frac{1}{8} (\mu p^+)^2 g_s B^2\cr
           &=\mu\cdot \frac{1}{8} g_2 B^2 =\mu\cdot \frac{1}{8 g_{eff}^2} B^2\ ,
\end{split}
\ee
where $g_2$ is the effective coupling for strings in the plane-wave background ({\it cf.} 
Appendix~\ref{effective-coupling}) and $g_{eff}\ (=1/\sqrt{g_2})$ is the effective coupling of 
the (gauge) theory residing on the giant graviton \cite{Hh}. 

This result is somehow what one would expect:   
The constant \bg $B$-field can also be understood as a constant magnetic field on the brane
(e.g. see \cite{SW}), and the energy, in units of $\mu$,  stored in a magnetic field $B$ in a 
gauge theory with coupling $g_{eff}$ is exactly the expression given in the second line of 
\eqref{V(0)}. 

In the absence of the $B$-field the potential felt by the radial fluctuation, which is obtained 
by inserting $X^i=R_0 r x^i$ into  \eqref{V}, and setting $B=0$:
\be\label{V(r)}
            V(r) = \frac{R_0^6}{2p^+ g_s^2}r^2(r^2-1)^2=\mu\cdot \frac{1}{2g_{eff}^2} r^2(r^2-1)^2 
\ee
has a maximum at $r^2=1/3$. 
(This is the potential studied in \cite{McGreevy:2000cw, Grisaru:2000zn, 
Hashimoto:2000zp}.) 
The value of the potential at this maximum is
\be\label{V(r)max}
            V^0_{max}=\mu\, \frac{2}{27g_{eff}^2} \,.
\ee
One would then expect that when $V^{(0)}$ becomes equal to $V^0_{max}$ or larger, the potential 
loses the minimum, so that the giant graviton becomes unstable and rolls down toward the minimum 
at $r=0$. This would happen for $B$-fields larger than the critical value
\be\label{B0c}
            B_{cr}^{(0)}=\frac{4}{3\sqrt{3}} \,.
\ee
The above estimate for $B_{cr}$ is a rough one and there are two points which should be 
taken into account. Adding the $B$-term  increases the energy at the maximum and also
changes the value of $r$ at which the potential is maximized. In fact there is a term 
proportional 
to $r^6B^2$ which should be added to \eqref{V(r)}. This effect would increase $B_{cr}$ from 
\eqref{B0c} level to about $\sqrt{{4}/{3}}$. This will be discussed in more detail in sections 
\ref{r-lambda-potential} and \ref{ASD}. 
Moreover, and as we will show momentarily, besides the resizing, shape of the giant can 
(will) change, lowering the ``vacuum'' energy 
from $V^{(0)}$ as well as increasing $V^0_{max}$.

The next term in \eqref{V} is the term linear in $Y^i$. This term is responsible for the 
resizing and also reshaping of the giant graviton. 
As discussed in \cite{Hh} and can also be readily seen from \eqref{V},  in the absence of the 
$B$-field this term vanishes and hence, similarly to the $V^{(0)}$ term, this term is 
second order in $B$.  It is straightforward to show that\footnote{Note that the 
``canonically normalized''
fluctuation is $\sqrt{\mu p^+} Y_i$, {\it cf.} discussions of section 2.4 of \cite{Hh}.}
\be\label{force1}
\begin{split}
            V_i^{(1)} Y^i & =\frac{3}{4p^+ g_s^2}\,  B^2 R_0^5 \, x_i Y^i  
                        -  \frac{R_0^5}{ p^+ g_s^2}\; T_{ij}(B) x^i Y^j \cr
&= \mu\ \frac{1}{4g_{eff}}\left(3B^2\delta_{ij}-4T_{ij}\right) x_i (\sqrt{\mu p^+} Y_j)\ .
\end{split}
\ee
The $V_i^{(1)} Y^i$ term of the potential (\ref{force1}) results in a force acting on the 
spherical brane which consists
of two components. The component proportional to $x_i Y^i $ is responsible for an overall 
resizing of the brane, whereas the component proportional to $T_{ij}(B)\, x^i Y^j$ causes 
changes in its shape. 

To find a new size and shape of the brane, we introduce an ansatz  with a corrected background,
\be\label{deform_ans}    
           X^i = R_0\, x^i + Y_0^i(B) + Z^i \,,
\ee
where $Y_0^i(B)$ is to be fixed requiring that there is no linear term in $Z^i$ in the 
expansion of the 
potential. Explicitly, $Y_0^i$ should satisfy
\be\label{Y0-full}
           V^{(1)}_i  + 2V^{(2)}_{ij}\, Y_0^j+3 V^{(3)}_{ijk}\, Y_0^jY_0^k +\ldots = 0  \,.
\ee
Assuming that the reshaping and resizing are small, {\it i.e.} $Y_0^i\ll R_0$, we can 
neglect higher 
order terms in \eqref{Y0-full}. Since the force term is proportional to the $B$-field, small 
$Y_0^i$ assumption is equivalent to a similar assumption on the $B$-field, $B^2\ll 1$, in which 
case
\eqref{Y0-full} reduces to
\be\label{Y1}
           V^{(1)}_i  + 2V^{(2)}_{ij}\, Y_0^j=0\ .
\ee

All the geometric fluctuations of a giant three-sphere have been analyzed in \cite{Hh}, and in 
particular it was noted that these fluctuations can be classified in terms of $SO(4)$ harmonics 
or, in other words, by irreducible representations of $SO(4)$. 
Moreover, it was  noted that $V^{(2)}_{ij}$ is diagonalized by $SO(4)$ spherical harmonics.
Hence $Y_0^i$ that solves \eqref{Y1} should have the same $SO(4)$ harmonic structure as the force 
term $V^{(1)}_i$, explicitly
\be\label{Y0_ans}
           Y_0^i(B) = R_0 \left[ (r(B)-1)\, x^i + \lambda S_{ij}(B)\,x^j\right] \,,
\ee
where $S_{ij}$ similarly to $T_{ij}$ is a symmetric traceless tensor of $SO(4)$, and 
$r$ and $\lambda$ are two variables which should be solved for (as functions of $B$).
Solving \eqref{Y1} we find
\be\label{rO}
\begin{split}
           S_{ij}&=T_{ij}\ ,\cr
           r(B) = 1 - \frac{3}{16} B^2+{\cal O}(B^4) \,&,      \quad     \lambda(B) = 
\frac{1}{4}+{\cal O}(B^2)\, .
\end{split}
\ee
As we see, the correction to the radius is negative. That is, the size of the giant graviton 
is reduced under the $B$-field, while its shape is ``squashed'' with the ``stress'' tensor
of the $B$-field, $T_{ij}(B)$. It is worth noting that the new shape and size do not depend 
on $g_{eff}$ and are only  functions of the $B$ field.

Before moving to a complete and general analysis of reshaping and resizing we would like to 
reconsider our zero point energy analysis.  
If the reshaped giant graviton is stable, along with our force arguments we expect that reshaping 
should decrease the zero point 
energy. The corrected zero point energy is then obtained as
\be\label{tildeV0}
            \tilde{V}^{(0)} = V^{(0)} - V^{(2)}_{ij}\, Y_0^i Y_0^j + \ldots  \,, 
\ee
which up to the fourth order in $B$ is
\be\label{zpe4}
\begin{split}
            \tilde{V}^{(0)} &= V^{(0)} - (V_0)^{(2)}_{ij}\, Y_0^i Y_0^j +{\cal O}(B^6)\cr 
                            &=\mu\cdot \frac{1}{8g_{eff}^2}\left[{B^2} - \frac{1}{16} 
                          \biggl(9(B^2)^2 + 4T^2(B) \biggr)   \right]+{\cal O}(B^6)\\ 
                        &=\mu\cdot \frac{1}{8g_{eff}^2} B^2\left(1-\frac{3}{16}B^2\right)^3
                             -\mu\cdot \frac{1}{32g_{eff}^2}T^2+{\cal O}(B^6)  \,,
\end{split}
\ee
where $T^2 = T_{ij}T^{ij}$. To perform the above computation we have used
the formula \cite{Hh}
\be\label{2nd}
 (V_0)^{(2)}_{ij}\, Y_0^i Y_0^j = \frac{\mu^2 p^+}{2} (Y_0^i +  \mathcal{L}_{ij}  Y_0^j)^2\,,
            \qquad \mathcal{L}_{ij}=x_j\partial_i-x_i\partial_j   \,,
\ee 
and the expression for $Y_0^i$ (\ref{Y0_ans}), (\ref{rO}).
It is worth noting that in the third line of \eqref{zpe4} the first term is  the energy 
stored in the magnetic field $B$ on a three-sphere of radius $R_0\,r$ (it encodes 
resizing of the giant graviton), whereas the second term, which is proportional to $T^2$,
comes from reshaping of the brane. In other words \eqref{zpe4} is the energy stored in a 
magnetic field $B$ on the squashed giant.

\subsection{Detailed analysis of the potential}\label{r-lambda-potential}

In the previous section we assumed that the reshaping and resizing are small, which is 
equivalent to a similar assumption on the $B$-field, $B^2\ll 1$. In the lowest order in the 
$B$-field,
we found that  reshaping of the brane is described by the lowest (linear in $x^i$) harmonics of 
$SO(4)$. In this section, we study the potential for arbitrary values of the $B$-field, and 
show that the reshaping is always described by the lowest harmonics of  $SO(4)$. 
In other words, the ansatz \eqref{Y0_ans} remains valid for the large $B$-field, and 
one does not have to include higher $SO(4)$ harmonics.
 
To study the effects of a large $B$-field, it is convenient to use the equations of motion
for the embedding $X^i = X^i(\sigma)$ that extremize the potential \eqref{V}. 
To obtain these equations we compute  variation of the potential $\delta V$ under the 
variations of the embedding $\delta X^i$, and equate it to zero. The result is
\bea\label{eom}
          2R_0^4\, X^i - \frac{4}{3}\,R_0^2\, \epsilon_{ijkl} \{ X^j , X^k , X^l \} 
                + \frac{1}{6}\, \epsilon_{ijkl}\epsilon_{jmnp}\{ \{ X^m , X^n , X^p \} , X^k , X^l \} \cr
                \cr
          - 2\{ \{ X^j , X^l , X^m \} , X^k , X^m \} \; B_{ij} B_{kl} 
                - \{ X^m , X^k , \{ X^j , X^l , X^i \}\} \; B_{mj} B_{kl}  = 0 \ .  
\eea

Now let's plug  the ansatz of the form
\be\label{higherS}
           X^i(x) = \sum_n \, S^i_{i_1 i_2 \ldots i_n}(B)\, x^{i_1} x^{i_2} \ldots x^{i_n}  \ ,
\ee
 into \eqref{eom} and note that except the first term in \eqref{eom} all the other terms, in 
particular the terms 
proportional to the $B$-field, only involve Nambu brackets of $X^i$'s.  
Using the Leibnitz rule for the Nambu bracket and equation \eqref{units3}
one observes that the equations \eqref{eom} reduce to a system of algebraic equations that 
relate the components $S^i_{i_1 i_2 \ldots i_n}$ of ranks $n$, $n' = \frac{n+2}{3}$, and 
$n'' = \frac{n+4}{5}$.  This system does not close for any finite number of terms in 
\eqref{higherS}, except for the trivial solution $X^i = 0$ and the linear solution with 
$n=1$.  These two 
solutions are in agreement with the results of the perturbative analysis of the previous 
section, where we found
the point-like solution and the squashed giant graviton solution
\eqref{deform_ans}, \eqref{Y0_ans}, \eqref{rO}. Since for a small $B$ we didn't see any solutions 
with infinitely many higher harmonics, which could in principle
solve \eqref{eom}, we do not expect them to appear for arbitrary values of  $B$. 
Therefore, the linear ansatz \eqref{Y0_ans} already includes all the harmonics needed for
arbitrary values of the $B$-field.  Then, noting \eqref{T^2ij},  
\be\label{gen_ans}
         X^i = R_0\, D_{ij}(B)\,x^j \,, \qquad
           D_{ij}(B) = r(B)\delta_{ij} +\lambda(B) T_{ij}(B)  \,,
\ee
is the most general ansatz which solves \eqref{eom}. 
(Indeed, using \eqref{T^2ij} it easy to see that 
all second rank tensors made out of higher powers of $B_{ij}$ are proportional to either 
$\delta_{ij}$ or $T_{ij}$.)  
Thus, for a given $B$-field the problem reduces to the study of 
the potential as a function of just two variables $r$ and $\lambda$. 

Using the (anti-)selfdual decomposition \eqref{Bpm} and formulas \eqref{T^2}, \eqref{B^2}, 
it is convenient to introduce the ``balance parameter'' 
\be\label{gamma}
        \gamma^2 = \frac{4 T^2}{(B^2)^2} = \frac{4 (B^+)^2 (B^-)^2 }{ [(B^+)^2 + (B^-)^2]^2}\ ,
\ee
in addition to  $B^2 = B_{ij}B^{ij}$, to describe the $B$-field background. This parameter 
ranges from zero to one, vanishes for 
an (anti-)selfdual $B$-field, and is equal to 1 for  a ``balanced'' ($B^+ = \pm B^-$) $B$-field.
Also, it is convenient to use the ``shape parameter'' 
\be\label{s}
         s = \frac{1}{4}\, \gamma \lambda B^2 
\ee 
instead of $\lambda$. The change of variables from $\lambda$ to $s$ would become ambiguous 
for $\gamma=0$, the (anti-)selfdual $B$-field case, which would be studied separately in 
section \ref{ASD}. In this section we only consider 
$\gamma\neq 0$. Plugging the 
ansatz \eqref{gen_ans} into the potential \eqref{V} we find
\bea\label{VTrcs}
        V =  \frac{R_0^6}{8p^+ g_s^2}\,
                [\Tr D^2 - 8\, \mbox{det}D + \frac{1}{ 6}((\Tr D^2)^3 - 3\,\Tr D^2\, \Tr D^4 
+ 2\,\Tr D^6)\cr
                +\Tr (D^2 B D^4 B) - \frac{1}{ 2}\Tr (D^2 B D^2 B)\,\Tr D^2] \ .
\eea
Making use of the formulas given in Appendix~\ref{BT} and introducing the parameters 
$\gamma$ and $s$ \eqref{gamma}, \eqref{s}, we obtain
\bea\label{Vrs} 
        V =  \frac{\mu}{8g_{eff}^2} \{4[ r^2 (r^2 - 1)^2 
            + s^2 (s^2 - 1)^2 + r^2 s^2 (4 - r^2 - s^2)] \cr
            + B^2\ (r^2 - s^2)^2 (r^2 + s^2 - 2 \gamma\  rs)\}  \ .
\eea
Note that all the dependence on  $g_{eff}$ has been factored out and hence the reshaping and 
resizing are independent of $g_{eff}$ and are only  functions of $B$-field.
This potential has an obvious symmetry 
\be\label{rs}
         r\leftrightarrow s \ , 
\ee
which in fact is a part of $SO(4)$ rotational symmetry of the problem. To see this, notice that 
the matrix $\mathcal{O}_{ij} = \frac{2}{ T}\ T_{ij}\ ,\  T=\sqrt{T^2}\ ,$ is orthogonal, 
$\mathcal{O}^T \mathcal{O} = \mathbb{I}$. In other words, 
\be
        X^i = \frac{2}{ T}\ T_{ij} x^j
\ee 
gives another embedding of a round $S^3$ into $\mathbb{R}^4$.
Using the ansatz \eqref{gen_ans} rewritten in the form 
\be\label{rs-ansatz}
        X^i(r, s) = R_0 (r\delta_{ij} + s \mathcal{O}_{ij}) x^j  \,,
\ee 
we find 
\be\label{sym}
        \mathcal{O}_{ij} X^j(r, s) =  X^i(s, r) \ . 
\ee
Since $\mathcal{O}_{ij}$ is a constant $SO(4)$ rotation matrix, 
\[
        V(X^i, X^a) = V(\mathcal{O}_{ij} X^j, X^a) \ .
\]

\begin{figure}
\centering
\epsfig{figure=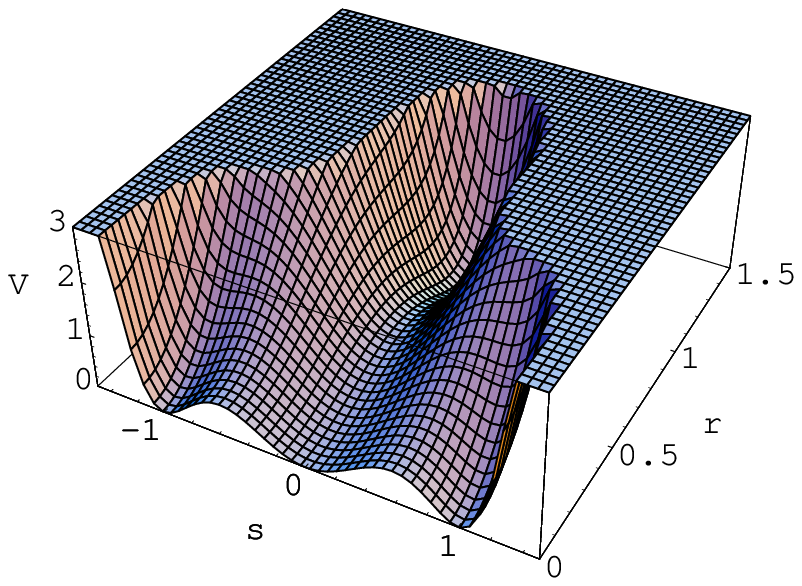} 
\begin{center}
\caption{Potential as a function of $r$ and $s$ for $\gamma=1$ and $B=0$.} 
\label{fig-B0}
\end{center}
\end{figure}

\begin{figure}
\centering
\epsfig{figure=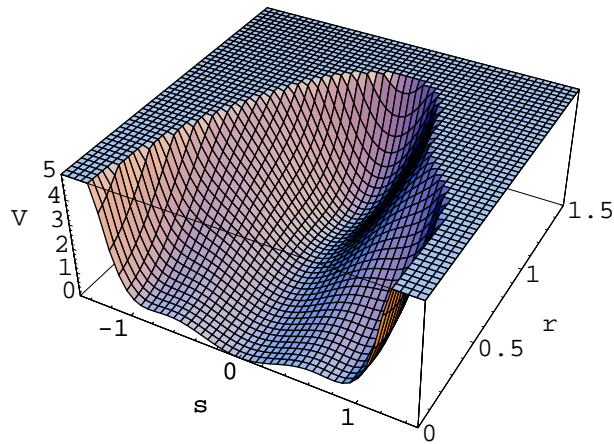} 
\begin{center}
\caption{Potential as a function of $r$ and $s$ for $\gamma=1$ and $B=1$.} 
\label{fig-B1}
\end{center}
\end{figure}

\begin{figure}
\centering
\epsfig{figure=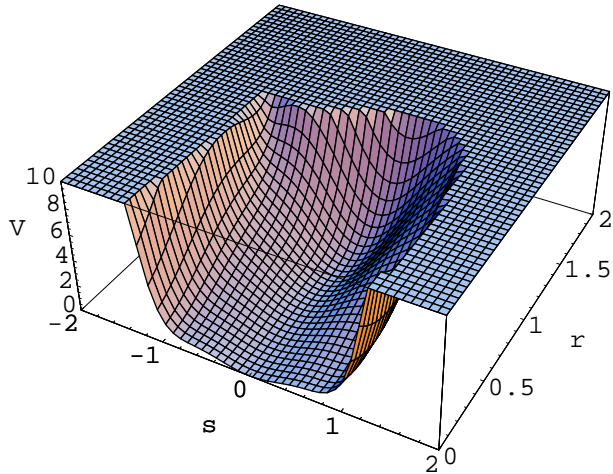} 
\begin{center}
\caption{Potential as a function of $r$ and $s$ for $\gamma=1$ and $B=1.5$.} 
\label{fig-B1_5}
\end{center}
\end{figure}

\begin{figure}
\centering
\epsfig{figure=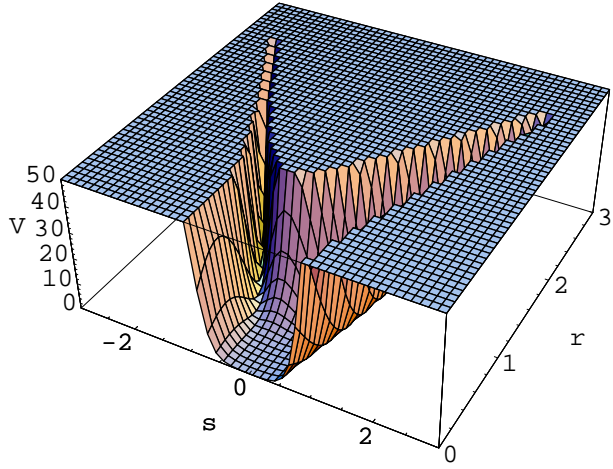} 
\begin{center}
\caption{Potential as a function of $r$ and $s$ for $\gamma=1$ and $B=10$.} 
\label{fig-B10}
\end{center}
\end{figure}

Profile of the potential \eqref{Vrs} for the case of the ``balanced'' $B$-field ($\gamma = 
1$) for different values of $B = \sqrt{B^2}$ is shown in Fig.~\ref{fig-B0}-\ref{fig-B10}. 
As we see in Fig.~\ref{fig-B0}, for the vanishing $B$-field we have minima at $s=0$ and 
$r=0$, $r=1$.
These are the point-like and spherical brane solutions studied in  \cite{McGreevy:2000cw}.
Note that the minima at $r=0$, $s=\pm 1$ describe the same spherical brane as 
the minimum at $s=0, \ r=1$, due to the symmetry \eqref{rs} and additional symmetry $r\leftrightarrow -s$
that appears at $B=0$. The plot in Fig.~\ref{fig-B1} corresponds to $B=1$ and shows the situation
when the minimum at $s=0, \ r\lesssim 1$ is lifted and is about to disappear, and the brane 
is about to roll toward 
$X=0$ vacuum. This can be easily seen on the $r=0$ section of the plot, given the 
symmetry \eqref{rs}. Using numerical analysis we found that the minimum disappears at 
$B_{cr}\approx 1.184$.
Fig.~\ref{fig-B1_5}, corresponding to $B=1.5$, shows the situation when the minimum at
$r\neq 0$ has already disappeared, so that the only minimum left is point-like $r=s=0$. One observes here
how large $B$ effects related to the second term in \eqref{Vrs} begin to dominate. The potential
starts developing two valleys at $r=\pm s$, which become dominant in the $B=10$ case shown 
in Fig.~\ref{fig-B10}.

\begin{figure}
\centering
\epsfig{figure=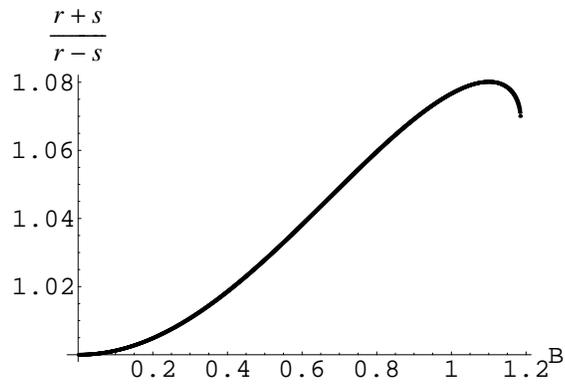} 
\begin{center}
\caption{Parameter $q_c=\frac{r+s}{r-s}$ as a function of $B$ for $\gamma=1$. At $B\approx 1.184$ 
the brane shrinks to a point, and the parameter $q_c$ can no longer be used to describe 
the shape of the brane.} 
\label{fig-q}
\end{center}
\end{figure}

It is also important to know how the shape of the brane depends on the value of $B$-field, 
before the brane
shrinks to a point. The shape of the brane is described by the ``shape parameter'' $s$ \eqref{s}, 
which is equal to zero for a round sphere solution. It turns out, however, that it is more 
convenient to introduce the parameter
\be\label{qq}
       q_c = \frac{r+s}{r-s}  \ ,
\ee 
which is equal to one for a round sphere. (The subscript $c$ on $q$ shows that this parameter 
measures the out-of-sphericity for the closed string metric.) The dependence of $q_c$ on $B$, 
for $\gamma=1$ $B$-field is depicted in Fig.~\ref{fig-q}.
One observes that $q_c(B)$ grows with $B$ monotonically almost everywhere, except the 
region of $B$ close to the critical value $B_{cr} \approx 1.184$, where $q_c$ decreases with 
$B$.

For the rank one $B$-field ($\gamma=1$) in the large $B$ limit, 
\be\label{limit_eff}
      B\rightarrow\infty\ , \quad  \frac{B^2}{g_{eff}^2} = const\ ,
\ee      
the first term in \eqref{Vrs} disappears, and the valleys $r=\pm s$ become flat directions. 
In other words, it costs no energy to roll in the directions $r=\pm s$. To find the shape of 
the brane in this case we note that the embedding \eqref{rs-ansatz} takes the form 
\be\label{r=s}
        X^i(r, s) = R_0 r (\delta_{ij} \pm \mathcal{O}_{ij}) x^j  \ .
\ee 
In the next section we show that $\mathcal{O}_{ij}$ can be brought into a diagonal form 
\eqref{Odiag}.
Taking ``$+$'' or ``$-$'' sign in \eqref{r=s}, one finds the embeddings
\bea\label{plane}
        X^{1, 2} = R_0 r \  x^{1, 2} \ , \qquad X^{3, 4} = 0 \ , \cr
        X^{3, 4} = R_0 r \  x^{3, 4} \ , \qquad X^{1, 2} = 0 \ ,
\eea
where now $r$ can take any arbitrary value and hence \eqref{plane} describes a plane which 
grows in the $X^{1, 2}$ or $X^{3, 4}$ directions.

Given the definition of $g_{eff}$ \eqref{V(0)}, the limit \eqref{limit_eff} is equivalent to
\be\label{limit_gs}
        B\rightarrow\infty\ , \quad  g_2 B^2 ={\rm  const} .
\ee
where $g_2$ is the  coupling for the closed strings in the plane-wave background.
In other words, in this limit $B$-field is large and the string coupling is small. In this  
regime the brane loses one of its dimensions and becomes a two dimensional plane. 

\subsection{Open string parameters}\label{open} 

Equation \eqref{rs-ansatz} (or \eqref{gen_ans}) is basically determining shape of 
the three-brane giant embedded in $\mathbb{R}^4$. Using these it is  straightforward to 
compute the 
induced metric on the brane
\be\label{induced-metric}
\begin{split}
        ds^2& = \frac{\partial X^k}{\partial x^i}\frac{\partial X^k}{\partial x^j}dx^idx^j\cr
                & = R_0^2\left[(r^2+s^2)\delta_{ij}+2rs{\cal O}_{ij}\right]dx^idx^j \ ,
\end{split}
\ee
where $r$ and $s$, which are functions of $B$, are minimizing the potential \eqref{Vrs}.
The metric \eqref{induced-metric} is in fact giving the shape of the brane viewed by 
closed strings, {\it i.e.} \eqref{induced-metric} is the closed string metric.

Due to the term proportional to ${\cal O}_{ij}$ for a generic $B$-field  $SO(4)$ rotational 
isometry of the giant is reduced to a $U(1)\times U(1)$ subgroup.
To see this note that we can always find a basis in which 
$B_{12}=-B_{21}$ and  $B_{34}=-B_{43}$ are the only non-zero components of the $B$-field.  
In this basis the two $U(1)$ are simply rotations in $12$ and $34$ planes and the non-vanishing 
components of $T_{ij}$ are
\be\label{Tij-components}
          T_{11}=T_{22}=\frac{1}{2}(B^2_{12}-B^2_{34})\ ,\ \ \ T_{33}=T_{44}=-T_{11}
\ee
and $T^2\equiv T_{ij}T^{ij}=4T_{11}^2$. Without loss of generality we take $T_{11}> 0$. 
Then, for the  $T\neq 0$ case, {\it i.e.} when $B\neq 0$ 
or $B_{12}\neq \pm B_{34}$,  
\be\label{Odiag}
          {\cal O}_{ij}=\mbox{diag}(1, 1,-1, -1) .
\ee
(The case $B_{12}= \pm B_{34}$ which correspond to (anti-)selfdual $B$-field will be considered 
separately in section \ref{ASD}.)

To work out the explicit form of the metric \eqref{induced-metric}, we adopt the coordinate system
\be\label{coordinate-systemI}
\begin{split}
x^1+ix^2\equiv z_1&=\cos\frac{\theta}{2}\ e^{i\alpha}\ ,\ \ \  0\leq \alpha,\beta \leq 2\pi, \cr
x^3+ix^4\equiv z_2&=\sin\frac{\theta}{2}\ e^{i\beta}\  ,\ \ \ 0\leq \theta\leq \pi,
\end{split}
\ee
in which the closed string metric takes the form 
\bea \label{csmetric-gen}
      ds^2&=& R_0^2\left[(r+s)^2 dz_1d{\bar z}_1+(r-s)^2 dz_2d{\bar z}_2\right]\\
             &=&\frac{R_0^2}{4}\left[(r^2+s^2-2rs\cos\theta)\  d\theta^2+ 
           4{(r+s)^2}\sin^2\frac{\theta}{2}\  d\alpha^2 + 4{(r-s)^2}\cos^2\frac{\theta}{2}\  
d\beta^2  \right]\nonumber
\eea

The embedding \eqref{coordinate-systemI} explicitly demonstrates the two $U(1)$ symmetries (as 
rotations in $z_1, z_2$ planes) and also the $r\leftrightarrow s$ symmetry. There is another 
$Z_2$ 
symmetry which exchanges $z_1$ and $z_2$ (together with $s \leftrightarrow -s$). These $Z_2$ 
symmetries are reminiscent of the original $SO(4)$. {}From the metric (\ref{csmetric-gen}) 
it 
becomes clear that $q_c$ defined in \eqref{qq} is indeed a measure of deformation of the 
round sphere.

By now it is well-known that D-branes in the constant $B$-field background, probed by the
open strings, behave as noncommutative surfaces \cite{{SW},{AAS}}, e.g. the low energy
effective theory residing on these branes is a noncommutative SYM theory \cite{ShJ}.
Moreover, the metric and the coupling viewed by open strings is different than those of
closed strings. In our case this implies that the shape of the giant graviton seen by open
strings is different than the one given through metric \eqref{csmetric-gen}. Note that in our
case, unlike the flat D-brane case, the Seiberg-Witten limit \cite{SW} does not lead to the
decoupling of bulk closed strings.\footnote{
For the flat D-branes in the $B$-field background, it is possible to take $\alpha'\to 0$ 
limit in such a way that the open string
metric (and hence the open string mass scale) and noncommutativity parameter 
$\Theta$ are held fixed while the closed strings of the bulk become very massive \cite{SW}.
Massless (supergravity) modes of the closed strings are also
decoupled because the closed string coupling is also sent to zero, while the open string
coupling is kept finite. In the plane-wave case, however, all the closed string modes, 
including the supergravity modes are massive (i.e. they have a non-zero
light-cone mass) \cite{review}. Moreover, all the physical modes of the giant three sphere
which correspond to geometric fluctuations of the giant, are massive \cite{Hh}. The mass
scale for both of the closed strings and the giant fluctuations (the open string modes) is 
$\mu$. Turning on the
$B$-field, as we will study in section \ref{spectrum}, in the range that we can still use 
the giant graviton description ($B<B_{cr}$) would not change the spectrum of the fluctuations 
of the giant very much. So, for the squashed giant we 
do not have a decoupling limit similarly to the
Seiberg-Witten case.
} Nevertheless, the notion of open string parameters is a useful one. 
  
In \cite{SW}, a prescription of calculating the  
open string metric $G_{rs}$ and the noncommutativity parameter $\Theta^{rs}$
in terms of closed string ones 
was introduced: \footnote{As was discussed in \cite{SW, Lambda-sym} open string parameters are not 
invariant under the $U(1)$ gauge transformation which rotates $B$-field into the $U(1)$ gauge field 
on the brane, $F$. In particular, the above prescription is in a gauge in which the background 
(magnetic) field on the brane, $F$, is set to zero.}
\begin{subequations}\label{open-close}
\begin{align}
         G^{rs}& =\frac{1}{2}\left[(g_c+b)^{-1} +(g_c-b)^{-1}\right]\\
         \Theta^{rs} &=\frac{1}{2}\left[(g_c+b)^{-1} - (g_c-b)^{-1}\right] \ ,
\end{align}
\end{subequations}
where $g_c$ is the induced closed string metric \eqref{csmetric-gen}, and 
\be
\begin{split}
   b &= B_{ij}\frac{\partial X^i}{\partial x^k}\frac{\partial X^j}{\partial x^l}dx^k\wedge dx^l\cr
    &= R_0^2\left[B_{12} (r+s)^2 idz_1\wedge d\bar z_1+B_{34} (r-s)^2 idz_2 \wedge d\bar z_2\right]\cr
    &= \frac{R_0^2}{4}\sin\theta \left[ B_{12} (r+s)^2 d\theta\wedge d\alpha+ B_{34} (r-s)^2 
    d\theta\wedge d\beta\right] \ .
\end{split}
\ee
Using the above formulas it is straightforward to work out open string parameters for the general 
rank two $B$-field. However, here we only present the explicit expressions for a rank one 
$B$-field, $B_{12}=0,\ B_{34}=B/\sqrt{2}:$\footnote{Note that rank one $B$-field corresponds to the 
balance parameter  $\gamma=1$. This can be readily seen from \eqref{Tij-components}.}
\begin{subequations}\label{open-close-rank1}
\begin{align}
ds^2_{open}&=\frac{R_0^2}{4}\left[G_{\theta\theta}\left(d\theta^2 +4\frac{(r-s)^2 
\cos^2\frac{\theta}{2}}
{r^2+s^2-2rs\cos\theta}d\beta^2\right) +4 (r+s)^2\sin^2\frac{\theta}{2}d\alpha^2
\right]\\
\Theta_{\theta\beta}&=-\frac{BR_0^2}{4\sqrt{2}}(r-s)^2 \sin\theta 
\left(1+\frac{B^2/2}{\frac{(r+s)^2}{(r-s)^2}+\cot^2\frac{\theta}{2}}\right) \ ,
\end{align}
\end{subequations}
where 
$G_{\theta\theta}=r^2+s^2-2rs\cos\theta+\frac{1}{2}(r-s)^2\sin^2\frac{\theta}{2}B^2$.
Note that indices on $\Theta$ are lowered and raised by the open string metric.


\subsection{(Anti-)Self-Dual $B$-field background}\label{ASD} 

For a generic $B_{ij}$, $T_{ij}$ is non-zero and hence, generically the shape of the giant graviton is 
deformed. However, if $B$-field is self-dual or anti-selfdual ({\it i.e.} $B$ is in $({\bf 3, 1})$ 
or $({\bf 1, 3})$ of $SU(2)\times SU(2)\simeq SO(4)$) $T_{ij}$ is identically zero,
and shape of the giant, viewed by closed strings, remains a round $S^3$ (while we still have 
resizing). This can  directly be seen from the force term \eqref{force1}. In this section we 
consider this particular $B$-field. 

This case can be analyzed by setting $s=0$ in the potential \eqref{Vrs}:
\be\label{Vasd}
        V_{ASD} = \mu\  \frac{1}{2g_{eff}^2} \left[ r^2 (r^2 - 1)^2 + \frac{B^2}{4} r^6\right] \ .
\ee
For all values of $B$ this potential has a minimum  at $r=0$. For values of  $B$-field
less than the critical value
\be\label{b-critic}
        B_{cr}=\sqrt{4/3} \ ,
\ee
this potential has another minimum at 
\be\label{r-min}
        r^2_{min}= \frac{2}{12+3B^2}(4+\sqrt{4-3B^2})=\frac{1}{2-\sqrt{1-\frac{3}{4}B^2}}
\ee
as well as a maximum. The value of the potential at this minimum is 
\be\label{Vmin}
       V_{min}=\mu \frac{B^2}{4g_{eff}^2}\cdot 
\left(\frac{1}{2-\sqrt{1-\frac{3}{4}B^2}}\right)^2\cdot
                     \frac{1}{1+\sqrt{1-\frac{3}{4}B^2}}\ .
\ee

\begin{figure}
\centering
\epsfig{figure=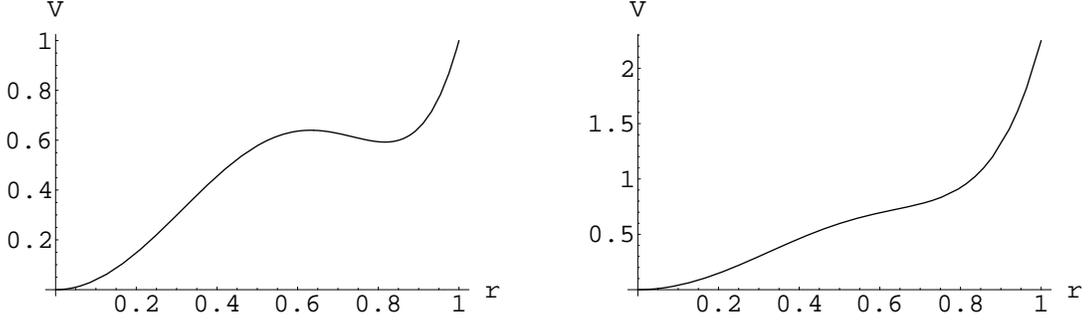} 
\begin{center}
\caption{Potential as a function of $r$ ($\gamma=0,\ s=0$) for $B=1$ (left) and $B=1.5$ (right).
The critical value of $B$-field is $B_{cr}=\sqrt{4/3} \approx 1.155$.} 
\label{fig-Vasd}
\end{center}
\end{figure}

This potential has been depicted in Fig.~\ref{fig-Vasd} for  values of the $B$-field below 
and above $B_{cr}$.
Note that the critical value of $B$-field where the giant graviton becomes classically unstable
depends on the ``balance parameter'' $\gamma$. As we discussed earlier, for $\gamma=1$ (rank one 
$B$-field) $B_{cr}\approx 1.184$ which is slightly more than in the selfdual case, $\sqrt{4/3}$. 
This can be understood simply by noting that the reshaping, on top of the resizing, would 
also 
decrease the energy (this can be seen from {\it e.g.} \eqref{zpe4}).

In the (anti)-selfdual case, although shape of the giant remains $SO(4)$ invariant, due to 
the 
background $B$-field this symmetry is reduced to a $SU(2)\times U(1)$. 
To see this,  let us start with an anti-selfdual $B$-field, $B_{12}=-B_{34}=B/2$.  
In the ``polar coordinates'' adopted in eq.~\eqref{SU2U1} of Appendix \ref{squasheds3} 
the pullback of this $B$-field on the round sphere of radius $R\equiv R_0 r_{min}$ is
\be\label{B-Asd}
b=\frac{B R^2}{8} \sin\theta d\theta\wedge d\phi .
\ee
As we see, $b$ is invariant under the $U(1)$ acting on $\psi$ coordinate, and the $SU(2)$ 
acting on 
the two sphere parameterized by $\theta,\phi$ ({\it cf.} Appendix \ref{squasheds3}).

This $SU(2)\times U(1)$ symmetry can also be explicitly seen in the open string parameters, and in 
particular open string metric. Working out the open string parameters using formulas of previous 
section we have
\begin{subequations}\label{openSU2U1}
\begin{align}
        ds^2_{open}&=\frac{R^2}{4}(1+B^2/4)\left[ (d\theta^2+\sin^2\theta d\phi^2)+ 
                 \frac{1}{1+B^2/4}(d\psi+\cos\theta d\phi)^2\right]\\
        \Theta_{\theta\phi} &= -\frac{BR^2}{8} (1 + \frac{B^2}{4}) \sin\theta \ .
\end{align}
\end{subequations}
The open string metric is a squashed sphere $S^3_q$ with the squashing parameter
({\it cf.} Appendix \ref{squasheds3})
\be
       q^2=\frac{1}{1+B^2/4}\ .
\ee
Note that $q\leq 1$.
The noncommutativity parameter $\Theta$, as we expect, is a constant two form on the $S^2$ base 
(it is proportional to the volume form of the base). It is worth noting that $\Theta$ is 
the flux of the magnetic field $B/2$ through the $S^2$ base of radius $R\sqrt{1+B^2/4}/2$.   

Upon quantization (of open strings) the $S^2$ base with the noncommutativity 
$\Theta_{\theta\phi}\propto \sin\theta$ becomes a fuzzy two sphere, $S^2_F$ {\it e.g.}  
see \cite{Ydri}. 
Consider the following embedding in a three dimensional noncommutative space with coordinates
$X^r$, $r=1,2,3$:
\be\label{fuzzys2}
        [X^r, X^s]=il\epsilon^{rsp}X_p\ , \qquad X_r^2=R^2 \ .
\ee
The fuzzy two sphere is described by a finite dimensional representation of the $SU(2)$ 
algebra whose generators are $X^r/l$; the radius of the sphere and the size of the matrices 
$N$ are related as \cite{DSV} 
\be\label{radiusN}
\left(\frac{R}{l}\right)^2=\frac{1}{4}(N^2-1) \ .
\ee
If we use the usual polar coordinates and write $X^r$ as functions of $\theta$ and $\phi$,
\eqref{fuzzys2} implies that 
\[
[\theta,\phi]=i \Theta_{\theta}^{\phi} \sim i\frac{1}{\sin\theta} \left(\frac{l}{R}\right) .
\] 
Comparing the above with (\ref{openSU2U1}b) we learn that the $B$-field should be quantized. For 
small $B$-field (or large $N$), i.e.
\be\label{quantized-B} 
          B=\frac{4}{N} \ .
\ee

\subsection{Reshaped giant as a giant bound state}

In the case of a flat D-brane, it was shown that \cite{Bound} a D$p$-brane in a constant
background rank one magnetic $B$-field which has both legs along the brane behaves as a
(non-marginal) bound state of $p$- and $(p-2)$- branes and the background $B$-field is giving
the density of the distribution of the RR charge corresponding to $(p-2)$-brane on the
D$p$-brane worldvolume.  If the worldvolume of D$p$-brane is along $012\cdots p$ directions
and the $B$-field is along $B_{p-1,p}$ then the corresponding D$(p-2)$-branes are along
$012\cdots p-2$ directions. {}From the string theory point of view, formation of this bound
state can be understood as follows. Although both of D$p$ and D$(p-2)$ -branes are
individually half BPS configurations, a system consisting of both of them is not.  This can
be seen from the fact that the open strings stretched between the a D$(p-2)$- and D$p$- brane
would become tachyonic once the separation between branes is of order of string scale (or
smaller). Due to the attractive force between the two branes formation of this tachyon is
inevitable and finally the two branes would become coincident, at which point the tachyon is
condensed and we end up with a (half) BPS brane, the D$p$-D$(p-2)$ brane bound state 
\cite{Bound}. The
process of ``dissolving'' of a $(p-2)$-brane into a $p$-brane, from the $p+1$ dimensional
gauge theory of $p$-brane viewpoint, is equivalent to turning on one unit of the magnetic
flux.

In the spherical brane and giant graviton case the story is somewhat different. Let us focus
on the three-sphere giants in the plane-wave background. Since there is no RR three form
flux in the background, a circular D-string is not  stable and the arguments made in the flat
D-brane case \cite{Bound} should be modified for this case. Moreover, spherical threebranes
and circular D-strings are not carrying a net RR charge, however, they are (electric) dipole
moments of the corresponding RR field.

Although the net RR charge of the giants is zero, locally they behave like a usual D-brane
and hence we expect the system of a three sphere giant in the background $B$-field to 
(locally)
behave like a bound state of a threebrane and D-strings. The simplest configuration of these
giant bound states is coming from the self-dual or (anti)-selfdual $B$-field case we studied
in the previous section. If the pullback of the $B$-field along the giant three sphere is
along a $S^2$ parametrized with $\theta, \phi$ then the {\it dissolved} circular D-strings
should be along $\sigma=d\psi\pm\cos\theta d\phi$ direction (the $\pm$ sign corresponds to
the self-dual or anti-self-dual $B$-field).  The density of the D-string RR dipole moment is
then proportional to $\Theta^{\theta\phi}$.

Dissolving of circular string giants into the three sphere giant can be seen from the
open string metric (\ref{openSU2U1}): Due to the tension of the circular D-strings the size
of the $S^1$ fiber along which they are wrapping is smaller than the $S^2$ base. Furthermore
note that the overall size of the giant due to this extra tension of the D-string giants has
been reduced (this e.g. can be seen from Figure 6).

\section{Spectrum of  fluctuations of the squashed giant}\label{spectrum}

Knowing the correct shape of the brane, one can find the spectrum of small fluctuations
around this configuration. In \cite{Hh}, the spectrum of small fluctuations around the
spherically shaped branes was computed. It includes both massive and massless (or zero)
modes. The zero modes are not physical modes and are gauge degrees of freedom, reminiscent 
of the area preserving diffeomorphisms on the three sphere giant, while the non-zero modes, 
which 
are of course all massive, correspond to the geometric fluctuations of the giant in the 
six directions transverse to the giant, \ie the $X^-$,  radial and $X^a$ directions. 
The mass of physical modes are all integer multiples of  $\mu$, a 
characteristic scale of the background, and are independent of the radius of the brane. 
In this section, we consider the (anti-)selfdual $B$-field case and compute the 
effects of the $B$-field on the spectrum of fluctuations of the 
squashed giant graviton in the $X^i$ and $X^a$ directions. In particular, as a confirmation 
of 
the spatial diffeomorphism invariance of the Born-Infeld action we show the existence of 
(three) zero modes. Furthermore, we find that the spectrum of the squashed giant, similarly 
to the unsquashed round case, is still independent of $p^+$ and $g_s$ and is only a function 
of the $B$-field.  
  

\subsection{Corrections to the spectrum of $X^i$ modes}\label{Xi-modes-1}

As we saw in the section~\ref{r-lambda-potential}, in an (anti-)selfdual background $B$-field 
the brane retains a spherical shape, changing only its size. In this case, the shape
parameter $s$ introduced in the section~\ref{r-lambda-potential}, is identically zero, and
the embedding of the brane \eqref{rs-ansatz} takes a form 
\be\label{s=0}
        X^i =  r(B)\ R_0 x^i \ .
\ee
The potential \eqref{Vrs} reduces to \eqref{Vasd}, and  $r(B)$ in \eqref{s=0} at the minimum of 
the potential must satisfy the condition
\be\label{extremum}
         1 - 4r_{min}^2 +3\left(1+ \frac{B^2}{4} \right)r_{min}^4 = 0\ ,
\ee
which has the solution \eqref{r-min}.
Let's parametrize small fluctuations around the embedding \eqref{s=0} by $Z^i$, {\it i.e.},  set 
\be\label{Z}
        X^i =  r(B)\ R_0 x^i  + Z^i \ .
\ee
As shown in \cite{Hh}, all fluctuations of the shape can be expanded in terms of $SO(4)$ 
spherical harmonics which are the eigenmodes of the operator $\mathcal{L}_{ij}=x_j\partial_i-x_i\partial_j$. 
We will assume that $Z^i$ in \eqref{Z} has an eigenvalue $\lambda$, explicitly
\be\label{LZ}
          \mathcal{L}_{ij} Z^j = \lambda \  Z^i  \ .
\ee
Using the definition of $\mathcal{L}_{ij}$  and the properties 
\eqref{units3}, one can rewrite \eqref{LZ} in the form
\be\label{xxZ}
         \epsilon_{ijkl} \{x^j , x^k, Z^l \} = -2\lambda Z^i  \ ,
\ee
or, equivalently, 
\be\label{3xxZ}
         \{x^j , x^k, Z^l \} + \{x^k, x^l, Z^j \} + \{x^l , x^j, Z^k \} = -\lambda \epsilon_{ijkl} \ Z^i  \ .
\ee
Plugging \eqref{Z} into \eqref{V} and using the formulas \eqref{xxZ} and  \eqref{3xxZ}, up to 
the second order in $Z$ we get 
\footnote{We also use here the integration by parts in the form $\int A\{B,C,D\} = - \int B\{A,C,D\}$.}
\be\label{Vzz}
       V =  \frac{\mu^2 p^+}{2}\ \left[1+4\lambda r^2 
               + r^4\lambda (\lambda - 2)\left(1+\frac{B^2}{4}\right)\right]\  Z^i Z^i  \ , 
\ee
where we have also used the fact that $T_{ij}(B) = 0$ for an (anti-)selfdual $B$-field.
Now, using the extremum condition \eqref{extremum}, we can rewrite \eqref{Vzz} as
\be\label{Vzz2}
       V =  \frac{\mu^2 p^+}{6}\ (\lambda + 1)\left[(4 r_{min}^2 - 1)\lambda + 3 
\right]\  Z^i Z^i  \ .
\ee
{}From this formula, noting that the kinetic term of the light-cone  Hamiltonian is 
$P^2/2p^+$, one can read off the frequency of the eigenmodes:
\be\label{omega-lambda}
      \omega^2_\lambda=\mu^2(\lambda+1)\left[\frac{\lambda}{3}(4r^2_{min}-1)+1\right]\ .
\ee
As discussed in \cite{Hh}, $\lambda$ can take three values: $\lambda = -1$,
\be\label{nz-modes1}
        \lambda = l\ , \quad l = 0,1,2, ... \ , 
\ee 
and
\be\label{nz-modes2}
        \lambda = -(l+2)\ , \quad l = 1,2, ... \ ,
\ee
where $l$ is the $SO(4)$ quantum number.
The modes with $\lambda=-1$ are zero modes and represent the non-physical gauge degrees of 
freedom.\footnote{The existence of  three zero modes is very important and is a cross 
check for the 
correctness of our calculations. One can then repeat the above calculations for the generic 
$B$-field. Inserting the expansion 
\[
        X^i =   D_{ij}(R_0 x^j  + Z^j) \ 
\]
into the potential, assuming that ${\cal L}_{ij}Z_j=-Z^i$ (the ``zero-modes'') and expanding 
to the second order in $Z$, we obtain
\[
        V^{(2)}|_{zero\ mode}=\frac{\mu^2 p^+}{2} Z_i D_{ik}\frac{\partial V}{\partial 
D_{kj}} Z_j
\]
where $V$ is given in \eqref{VTrcs}. It is then evident that for the squashed giant 
configuration, where ${\partial V}/{\partial D_{ij}}$ vanishes, $Z_i$'s satisfying ${\cal 
L}_{ij}Z_j=-Z_i$ are zero modes.}  
There are two sets of physical modes corresponding to the values of $\lambda$ 
\eqref{nz-modes1} and \eqref{nz-modes2}. 
Noting that $0\le B < B_{cr}=\sqrt{4/3}$ which leads to $\frac{1}{2} < r^2_{min} \le 1$ or
$1 < 4 r_{min}^2 - 1 \le 3 $, it is easy to 
see that all the non-zero modes for all values of $l$ have positive $\omega^2$, and 
therefore there are 
no tachyons in the spectrum. 
It is worth noting that the frequency of the $l=0$ mode, which 
corresponds to the center of mass motion of the squashed giant without changing its shape, is 
independent of the $B$-field and squashing.

For the $B=0$ case ($r_{min}=1$) both of the modes with $\lambda=l$ and  $\lambda=-(l+2)$ 
have the same mass $\mu (l+1)$ \cite{Hh}. This degeneracy is, however, lifted
due to the resizing corrections, and these two modes have now different masses.

For the critical $B$-field where $r^2_{min}=1/2$, the $l=1,\ \lambda =-3$ mode 
becomes massless while all the other modes  still have a positive mass squared.
This is a sign that the squashed giant would become unstable for $B> B_{cr}$. Furthermore,
this shows that the radial (breathing) mode is the mode along which the squashed giant 
develops instability.

Before moving to the other modes, we would like to point out that the above ``small 
fluctuation''   expansion  may break down for the modes with high $l$. More precisely, the 
above expansion can be trusted for the modes for which 
\[
        \omega_l \lesssim \mu\frac{B^2_{cr}-B^2}{g^2_{eff}} .
\]
In terms of the $SO(4)$ quantum number $l$, that is $l\lesssim 
\frac{3(B^2_{cr}-B^2)}{(4r^2_{min}-1) 
g^2_{eff}}$.  
 
%


\subsection{Corrections to the spectrum of $X^a$ modes}\label{Xa-modes-1}

Analogously to the $X^i$ modes, corrections to the spectrum of $X^a$ modes can be computed.
Consider small fluctuations around the solution $X^i = R_0 r \  x^i$, $X^a = 0$, {\it i.e.}, plug 
$X^i = R_0 r \  x^i$, $X^a = Z^a$ into the potential \eqref{potential}, \eqref{det} and keep only
terms of the second order in $Z^a$:
\bea\label{VZa1}
       V = \frac{\mu^2 p^+}{2}\left[Z^a Z^a + \frac{r_{min}^4}{2}\left( 
              \{ x^i , x^j , Z^a \} \{ x^i , x^j , Z^a \}  
          + \{ x^i , x^k , Z^a \} \{ x^j , x^l , Z^a \} B_{ij} B_{kl}\right)\right] .
\eea
To compute this expression, we go to the basis where an anti-selfdual $B$-field has the only
non-zero components $B_{12}=-B_{34}=B/2$, and assume that $Z^a$ are $SO(4)$ harmonics
with quantum numbers $(l, m_1, m_2)$ defined as
\footnote{Note that these numbers must be the same for all components $Z^a$, 
since the rotations in the $X^i$ and $X^a$ directions commute with each other. Note also 
that with our definition ${\cal L}_{ij}$ is anti-hermitian and hence eigenvalues of 
${\cal L}^2$ are negative and eigenvalues of ${\cal L}_{12}$ and ${\cal L}_{34}$ are 
imaginary.} 
\be\label{Casimir}
        \mathcal{L}_{ij}\mathcal{L}_{ij} Z^a = -2l(l+2)\ Z^a \ ,
\ee
\be\label{U1s}
        \mathcal{L}_{12} Z^a = i m_1\  Z^a \ , \quad  \mathcal{L}_{34} Z^a = i m_2\  Z^a \ ,
\ee
where $ -l \leq m_1, m_2 \leq l$. 
Now, using the formulas
\begin{subequations}\label{xxF}
\begin{align}
       \{ x^i , x^j , \Phi \} &= -\half\  \epsilon^{ijkl}\mathcal{L}_{kl} \Phi \, \\ 
       \epsilon^{ijkl}\mathcal{L}_{ij}\mathcal{L}_{kl} \Phi & = 0 \ ,
\end{align}
\end{subequations}
(the identity (\ref{xxF}b) can be explicitly verified using definition of ${\cal L}_{ij}$) 
and 
integrating by parts, we get
\be\label{VZa2}
       V=\frac{\mu^2 p^+}{2}\  
\left[ 1 + r_{min}^4 \left( l(l+2) + \frac{B^2}{4}(m_1-m_2)^2 \right)  \right] Z^a Z^a \ .        
\ee
Again, this expression is positive for all $(l, m_1, m_2)$, and there are no tachyons. 
The frequencies (masses) are
\be\label{omega-lmm} 
        \omega^2_{l m_1 m_2} = \mu^2 \left[ 1 + r_{min}^4 \left( l(l+2) 
              + \frac{B^2}{4}(m_1-m_2)^2 \right)  \right]  \ .
\ee
This should be contrasted with the masses of the $X^i$ fluctuations which have only $l$ 
dependence. 
Note also that the azimuthal dependence of the frequencies only appears in the combination
$m = m_1 - m_2$, and not $m_1$ and $m_2$ individually. This is related to the fact that
in the self-dual $B$-field case we remain with a $SU(2)\times U(1)$ symmetry.

It is straightforward to compute the masses for the general 
$B_{12}\neq \pm B_{34}$ case. In this case, however, one should note that 
besides 
the resizing parameter $r$, the frequencies also depend on the reshaping parameter $s$. 
Performing the calculations we obtain  
\be\label{omega-general}
\begin{split}
\omega^2_{lm_1m_2}=\mu^2 \biggl[1+& l(l+2)(r^2-s^2)^2+4rs\left((r+s)^2 
m_2^2-(r-s)^2 m_1^2\right)\cr
+& \left((r+s)^2 B_{12} m_2 + (r-s)^2 B_{34} m_1\right)^2\biggr].
\end{split}
\ee
The above expression reproduces \eqref{omega-lmm} for $s=0$ and $B_{12}=-B_{34}$.
It is interesting to note that in the limit \eqref{limit_eff} where the minimum is at $r=\pm 
s$, the spectrum \eqref{omega-general} becomes $l$ independent, moreover the spectrum 
would only depend on $m_1$ or $m_2$ (depending on the $+/-$ sign). This is compatible with 
the arguments at the end of section~2.3.
%

\section{Stability of the squashed giant}\label{stability}

In previous sections we studied deformation of the shape and corrections to the spectrum of a 
giant graviton due to the presence of a constant background NSNS $B$-field.
In this section we address  the stability of the squashed giant. As mentioned in previous 
section, there are no tachyonic modes in the spectrum of the geometric fluctuations of the
squashed giant. This implies the classical stability of the squashed giant. Of course one 
should 
remember that even classically the giant is not stable under the fluctuations whose 
energy 
are comparable to $\frac{B_{cr}^2-B^2}{g^2_{eff}}$.

In section~\ref{susy} we analyze the supersymmetry of the squashed giant, and in 
section~\ref{tunnelling} we study instability of the squashed giant under quantum (tunneling) 
effects.

\subsection{SUSY analysis}\label{susy}

To study the supersymmetry of the squashed giant, we first start with the supersymmetry
algebra of the background plane-wave. The plane-wave \eqref{background} is a maximally 
supersymmetric background with 32 supercharges half of which are kinematical
anti-commuting to the light-cone momentum $p^+$,  and the other half are dynamical which 
anti-commute to 
the light-cone Hamiltonian, explicitly
\[
\begin{split}
[P^+, Q]=0\ , &\quad [P^+, Q^\dagger]=0\cr
[H, Q]=0\ , &\quad [H, Q^\dagger]=0\ .
\end{split}
\]
\be\label{susy-algebra}
\begin{split}
\{Q_{\alpha\dot\beta}, Q^{\dagger\rho\dot\lambda}\} & =\delta_\alpha^\rho\
\delta_{\dot\beta}^{\dot\lambda}\ {\bf {H}}
+2\mu (i\sigma^{ij})_{\alpha}^{\rho}
\delta_{\dot\beta}^{\dot\lambda}\ {{{\bf J_{ij}}}}
+2\mu (i\sigma^{ab})_{\dot\beta}^{\dot\lambda} \delta_{\alpha}^{\rho}\ {{{\bf J_{ab}}}}\cr
\{Q_{\dot\alpha\beta}, Q^{\dagger\dot\rho\lambda}\}& =\delta_{\dot\alpha}^{\dot\rho}\
\delta_{\beta}^{\lambda}\ {\bf {H}}
+2\mu (i\sigma^{ij})_{\dot\alpha}^{\dot\rho}
\delta_{\beta}^{\lambda}\ {{{\bf J_{ij}}}}
+2\mu (i\sigma^{ab})_{\beta}^{\lambda} \delta_{\dot\alpha}^{\dot\rho}\ {{{\bf J_{ab}}}}
\end{split}
\ee
where $Q$'s are dynamical supercharges and the two indices on them are  Weyl 
index of the two $SO(4)$'s acting on the $X^i$ and $X^a$ directions. The above superalgebra 
can 
be identified as 
$PSU(2|2)\times PSU(2|2)\times U(1)_- \times U(1)_+$ where the $U(1)_\pm$ are
translation along the $x^\pm$ directions, generators of which are $p^+, H_{l.c}$. More 
detailed discussion on this superalgebra can be found in \cite{review}. Adding the $B$-field 
does not change the supergravity background and hence its superalgebra.

In the absence of the $B$-field a round three sphere giant is half supersymmetric, and it 
preserves all the dynamical supercharges. This can 
be readily  seen from the fact that this solution is a zero energy solution ($H=0$) and has 
zero charges under both of the $SO(4)$'s, i.e. it has $J_{ij}, J_{ab}=0$. Therefore, 
acting on the round giant graviton state,
the right-hand side of \eqref{susy-algebra}  vanishes. 

Fluctuations of the round giant are generically less supersymmetric  states.
In particular let us focus on the two modes which appear in the reshaping and resizing.
In the notations of \cite{Hh} both of these have $l=1$ under one of the $SO(4)$'s and have 
zero charge under the other $SO(4)$. The resizing mode (the ``breathing'' mode), $\delta X^i 
= x^i$ has zero $J_{ij}$ and $J_{ab}$ charge with energy $H=2\mu$. Hence this mode by itself 
is not  BPS at all (acting by this state, the right-hand-side of \eqref{susy-algebra} is 
non-vanishing).  The reshaping mode  
$\delta X^i= S_{ij}x^j$ which comes with degeneracy $3\times 3=9$ also has energy $2\mu$ and 
$J_{ab}=0$ whereas its $J_{ij}$ eigenvalues can be  $0$ or $\pm 1$. Therefore among the nine 
modes of the reshaping mode there are two modes which kill the right-hand-side of 
\eqref{susy-algebra} for half of the dynamical supercharges, i.e. they are 1/4 BPS.
(For a similar argument for $SU(4|2)$ superalgebra see \cite{DSV}.) 
It is straightforward to check that the reshaping and resizing modes fall into the same 
supermultiplet of the $PSU(2|2)\times PSU(2|2)\times U(1)_-$ superalgebra which contains a 
1/4 BPS state \cite{Hh}. In other words this multiplet is a 1/4 BPS (short) multiplet.

Now let us consider the squashed giant. Although the state of squashed giant does not kill 
the right-hand-side of the SUSY algebra \eqref{susy-algebra}, as we argued above, the 
deformation of the giant
from a round three sphere  can be described in terms of turning on a 1/4 BPS multiplet. 
In this sense the squashed giant is 1/4 BPS. This should be contrasted with the flat space 
case, where a D-brane in the background $B$-field preserves the same amount of supersymmetry 
as a usual D-brane, i.e. half BPS. 
Being 1/4 BPS one would expect that the shape of the squashed giant, at least perturbatively, 
should be stable (protected) under small geometric fluctuations of the giant. However, there 
is a subtlety which despite of being BPS
might make the shape unstable: in principle it is 
possible that some short multiplets combine and form a long (non-BPS) multiplet and hence 
receive corrections. Based on similar analysis which was done for a spherical membrane in the 
eleven dimensional plane-wave background \cite{DSV}, however, we expect the specific modes 
involved in the reshaping of the giant to be stable. Indeed this expectation is well 
supported noting the potential \eqref{VTrcs} and the fact that the $g_{eff}$ dependence of 
the potential is only in an overall factor. Therefore the value of $(r,s)$ which  minimize 
the potential, and hence the shape and size of the giant, is independent of $g_{eff}$. In 
other words, the shape of the brane is stable under perturbative corrections about 
$g_{eff}=0$. This, however, does not exclude non-perturbative instabilities of the giant.

\subsection{Quantum instability of the squashed giant}\label{tunnelling}

Supersymmetry considerations imply that the shape of the squashed giant should be 
perturbatively stable. However,  as one can explicitly see from \eqref{Vrs} and the potentials 
depicted in section 2, the energy of the squashed giant is non-zero \cf \eqref{zpe4} while 
the minimum at $X=0$ has always zero energy. This in particular means that the squashed 
giant can tunnel to the X=0 vacuum. To have an estimate of the 
tunneling rate let us focus on the (anti-)selfdual $B$-field case and the potential
\eqref{Vasd}, depicted in Fig.~\ref{fig-Vasd}. 

Using the WKB approximation the tunneling probability is equal to the negative of 
the exponential of the area encapsulated between the potential $V_{ASD}$ and line 
$r=r_{min}$:
\be\label{WKB}
P_{tunnel}\approx e^{-\frac{B^2_{cr}-B^2}{g_{eff}^2}\Delta r(B)}
\ee
where $\Delta r(B) = r_0-r_{min}$ with $V(r_0)=V_{min}$ and $\Delta r(B_{cr})=0$.
The dependence of tunneling probability on $g_{eff}$ is like $e^{-1/g_{eff}^2}$ and 
hence from the giant graviton (gauge theory)  viewpoint this tunneling is an instanton 
effect. Therefore the squashed giant is metastable and through instanton (non-perturbative) 
effects would tunnel into the stable $X=0$ vacuum where it decays into the supergravity modes.

\section{Discussion}

In this paper we discussed giant gravitons in the ten dimensional type IIB plane-wave 
background when a constant NSNS $B$-field is turned on along the giant. As we argued if the 
$B$-field has only one of its legs along the three-sphere giant it is not essentially 
felt by 
the giant (similarly to the flat D-brane case \cite{SW}.)
Moreover,  the shape of the giant would change as a result of the existence of the 
$B$-field so that generically we have a ``squashed giant''. 

As we showed in section 2, there is a critical $B$-field, $B_{cr}$, above which the squashed 
giant becomes classically unstable. This means that if we tune up the $B$-field from zero we 
start 
squashing (resizing and reshaping) the giant and at $B>B_{cr}$ the squashed giant  
configuration is no longer minimizing the potential. Hence in the $B>B_{cr}$ 
background 
the only minimum is at $X^i=X^a=0$ (the zero size brane). 
As argued in \cite{McGreevy:2000cw, Hh} this $X=0$ solution cannot be stable as a quantum 
mechanical vacuum of the theory and it is not clear yet what this vacuum would look 
like quantum mechanically.

As we discussed and can be seen in Figs. 1, 2 and 6 the potential has a maximum and one would 
wonder whether it is possible to expand the theory around the maximum of the potential where 
we have a system with open string  tachyons. The  giant graviton state is then 
where the tachyon is condensed (of course there is also the possibility that the tachyon 
rolls 
towards the $X=0$ vacuum). Although the above argument is generic for $B=0$ or $B\neq 0$, the 
non-zero $B$-field case has its own special and interesting features.  Let us focus on the 
selfdual $B$-field case. As it is seen from Fig. 6, it is possible to consider the 
interesting limit of  $B\to B_{cr}\ (B<B_{cr})$ and $g_{eff}\to 0$ while 
$(B^2_{cr}-B^2)/g_{eff}^2$ is kept fixed. In this limit the difference between the energies 
of the squashed giant and the $X=0$  vacuum is sent to infinity while keeping the energy 
difference between the squashed giant minimum and the maximum of the potential finite.
In this limit almost all the modes of the geometric fluctuations of the giant are also 
decoupled (note that $g_{eff}\to 0$) and we remain with the $l=1$ modes. The potential in 
Fig. 6  in this limit resembles that of a $c=1$ matrix model \cite{MV}. It is then very 
plausible to expect that the squashed giant system would provide us with a laboratory to 
study open string tachyon condensation. 

Here we mainly focused on the squashed giants in the plane-wave background, one might pose 
the same problem in the $AdS_5\times S^5$ background, for which we again expect to see a 
similar squashing behaviour. 

The other interesting open question is: what is the ${\cal N}=4$ gauge theory 
operator which is dual to the squashed giant? A 
simple proposal, based on the results we obtained here and the discussions of 
\cite{Balasubramanian:2001nh,Hh}, is that this operator is 
a subdeterminant type operator \cite{Balasubramanian:2001nh} in the BMN sector 
\cite{review, BMN} which has the 
appropriate $SO(4)\times SO(4)$ charges. More specifically, this operator should be obtained 
by the insertion of the covariant derivative 
of the gauge theory on $R\times S^3$, ${\cal D}_i$, 
in the combinations $T_{ij}=F^2_{ij}-\delta_{ij} F^2/4$ where $F_{ij}=[{\cal D}_i, {\cal 
D}_j]$ (this part would give the reshaping) and $F^2$ terms (for resizing), into the 
subdeterminant operators. Working out the explicit form of this operator is an interesting 
open question we postpone to future works. Once we have this operator one might then compute 
the decay rate of the squashed giant, an approximation of which is given in \eqref{WKB}, 
from the dual gauge theory viewpoint.   

Another interesting question which we briefly discussed is quantization of the squashed giant 
and also the (noncommutative) gauge theory living on the giant. This would generically lead 
to a quantization of the background $B$-field (\cf \eqref{quantized-B}). This direction 
deserves a more thorough and detailed analysis.

A problem similar to what we considered here for three-sphere giants may be asked about 
the spherical M5-branes in the eleven dimensional plane-wave \cite{DSV, BMN} or 
$AdS_{4,7}\times S^{7,4}$ background \cite{McGreevy:2000cw}. In analogy with our 
results for the three-sphere case, we expect 
in response to a background constant three form field 
the five sphere giant to be deformed (squashed), moreover  we expect this ``squashed'' five 
sphere giant 
to be a bound state of spherical M2-brane and spherical M5-brane giants. The theory residing 
on the ``squashed'' five sphere giant is then a deformation of the $(0,2)$ theory on $R\times 
S^5$. It would be nice to check if this deformation would provide us with an expansion 
parameter which could be used to make a perturbative analysis of the $(0,2)$ theory.


{\large{\bf Acknowledgements}}

We would like to thank Simeon Hellerman, Keshav Dasgupta, John McGreevy, Darius Sadri and 
Matt Strassler for fruitful discussions and Michal Fabinger for his collaboration at the 
early stages of this work.
The work of M. M. Sh-J. is supported in part by NSF grant
PHY-9870115 and in part by funds from the Stanford Institute for Theoretical
Physics. The work of S.P  is supported by NSF grant PHY-0244725.

\appendix

\section{Some useful identities}\label{BT}

The B-field, $B_{ij}\ i,j=1,2,3,4 $, we have been working with is in ${\bf 6}$ of $SO(4)$. In terms 
of $SU(2)\times SU(2)$ representations, however, this is a reducible one. That is, 
we can decompose $B$ into its self-dual and anti-self-dual parts, $B^+$ and $B^-$ respectively: 
\be\label{Bpm}
        B^+_{ij}=\frac{1}{2}(B_{ij}+\frac{1}{2}\epsilon_{ijkl}B_{kl}),\qquad  
        B^-_{ij}=\frac{1}{2}(B_{ij}-\frac{1}{2}\epsilon_{ijkl}B_{kl}) \,.
\ee
$B^+$ and $B^-$ are then in ${\bf (3,1)}$ and ${\bf (1,3)}$ of $SU(2)\times SU(2)$.

The ``energy-momentum'' tensor $T_{ij}$,
\[
        T_{ij}=B_{ik}B_{kj}+\frac{1}{4}\delta_{ij} B^2,
\]
which is in ${\bf (3,3)}$ of $SU(2)\times SU(2)$, takes a simple form once $B^\pm$ are used:
\be\label{TBpm}
        T_{ij}=2B^+_{ik}B^-_{kj}=2B^-_{ik}B^+_{kj}\ .
\ee
Using \eqref{TBpm} and the fact that $B^\pm_{ik}B^\pm_{jk}=\frac{1}{4}B^2\delta_{ij}$
we obtain a very useful identity
\be\label{T^2ij}
        T_{ik}T_{jk}=\frac{1}{4} T^2\delta_{ij}\ ,
\ee
where
\be\label{T^2}
        T^2\equiv T_{ij}T_{ij}= (B^+)^2 (B^-)^2  \,.
\ee
Noting that 
\be\label{B^2}
        B^2 = (B^+)^2 + (B^-)^2
\ee
and the above, $(B^+)^2$ and $(B^-)^2$ are then solutions of the 
quadratic equation $X^2-B^2 X+T^2=0$.

Let us define  the matrix $D_{ij}$,
\be
        D_{ij}=r\delta_{ij} +\lambda T_{ij}\ ,
\ee
where $r$ and $\lambda$ are two arbitrary $c$-numbers.
It is straightforward to show that
\be
\begin{split}
               \Tr D&=4r\  , \qquad \Tr D^2=4r^2+\lambda^2 T^2\ , \cr
               \Tr D^3 &= 3r\Tr D^2-8r^3=4r^3+3r\lambda^2 T^2\ , \cr
               \Tr D^4&=\frac{1}{4}\left(\Tr D^2\right)^2 + 4r^2 \Tr D^2 -16 r^4\ , \cr 
\Tr D^6 &=\Tr D^2\left[\frac{1}{16}\left(\Tr D^2\right)^2 + 3r^2 \Tr D^2 -12 r^4\right] \ ,\cr
\det D & =\frac{1}{16}(\Tr D^2-8 r^2)^2=\frac{1}{16}(4r^2-T^2\lambda^2)^2\ , 
\end{split}
\ee
\be
B_{ik}T_{kj}B_{ji}= T^2\ , \quad
T_{ik}B_{kj} T_{jl} B_{li}=-\frac{1}{4} B^2 T^2\ ,
\ee
and
\be
\begin{split}
        \Tr (D^n B)&=0, \quad n=0,1,2, \ldots\ , \cr
        \Tr(D^2 B D^2 B)& = - B^2\left[\frac{1}{16} \left( \Tr D^2\right)^2+r^2 \Tr D^2 -4r^4\right]
            +r\lambda T^2 \Tr D^2\ ,\cr
        \Tr(D^4B D^2 B)&=-B^2 \Tr D^2 \left[\frac{1}{64} \left(\Tr D^2\right)^2+\frac{3}{4}r^2 \Tr D^2 
            -3r^4\right]\cr &+r\lambda T^2 \left[\frac{3}{8}\left(\Tr D^2\right)^2+2r^2 \Tr D^2-8r^4\right] \ .
\end{split}
\ee

\section{Squashed three sphere}\label{squasheds3}

A round three-sphere has $SO(4)\simeq SU(2)_L\times SU(2)_R$ symmetry (isometries), however,  
only a part of this symmetry group can be made explicit in the metric once a coordinate 
system is adopted. For example in the coordinate system in which the line element is 
$R^2(d\theta^2+\sin^2\theta d\Omega^2_2)$, $SU(2)_D$, i.e. the diagonal part of 
$SU(2)_L$ and $SU(2)_R$ is manifest. It is possible to adopt another coordinate system in which
$SU(2)_L\times U(1)$ is explicit:
\be\label{SU2U1}
\begin{split}
z_1&= R\cos\frac{\theta}{2}\ e^{i(\phi+\psi)/2}\cr
z_2&= R\sin\frac{\theta}{2}\ e^{i(-\phi+\psi)/2}
\end{split}
\ee
with $0\leq \theta\leq \pi$, $0\leq \phi\leq 2\pi$ and $0\leq \psi\leq 4\pi$, 
which gives the embedding 
of a $S^3$ in $\mathbb{C}^2$. $(z_1\ \ z_2)$ behave like a doublet under $SU(2)_L$, and $U(1)$ 
is the translation along $\psi$ direction. In this coordinate system the line element on $S^3$ is
\be\label{three-sphere-metric}
d\Omega^2_3=\frac{R^2}{4}\left[(d\theta^2+\sin^2\theta d\phi^2)+(d\psi+\cos\theta d\phi)^2\right] \ .
\ee
In this coordinate system 
$S^3$ is realized as $U(1)$ fiber over an $S^2$ base. The $S^2$ base and $S^1$ fiber
have the same radii equal to $R/2$.

Squashed three sphere, $S^3_q$ is a deformation of round $S^3$ with $SU(2)\times U(1)$ 
isometry. The metric for the squashed 
three sphere can be written as
\be\label{squashed-three-sphere-metric}
d\Omega^2_q=\frac{R_q^2}{4}\left[(d\theta^2+\sin^2\theta d\phi^2)+q^{2} (d\psi+\cos\theta 
d\phi)^2\right] \ ,
\ee
where $q$ is the deformation (squashing) parameter and may be taken smaller or larger than 
one.

\section{Effective closed string coupling in the plane-waves}\label{effective-coupling}

It is well-known that upon compactification of strings the effective coupling, 
i.e. effective Newton constant, for strings in lower dimensions and the original 
uncompactified theories are related by a factor of $\sqrt{V}$, where $V$ is the volume of the 
compactification manifold, in such a way that for infinite $V$ case the coupling of the lower 
dimensional theory is vanishing (note that this is only true when there is no warping factor in 
the non-compact part). For the case of strings on the plane-wave, however, because of the 
``harmonic oscillator potential'' (arising from the $(dx^+)^2$ term in metric once the light-cone 
gauge is employed) in a sense the situation is very similar to a compactification on an eight dimensional 
manifold \cite{G7}. To see this more concretely, let us consider a scalar field theory on the ten 
dimensional plane-wave \cite{Smat}. The classical equations of  the scalar field, once the 
interaction terms are turned off, can be exactly solved and the dependence of the wavefunction on 
the transverse coordinates is given in terms of harmonic oscillator wavefunctions, or Hermit 
polynomials while it is like a free particle moving in the $x^+$ and $x^-$ directions \cite{Smat}.
Therefore, the particles in the plane-wave background can only freely move along $x^{\pm}$ 
directions and (depending on their light-cone momentum $p^+$) they have only access to a finite 
volume in the transverse directions. (One may then consider interactions, and analyze the theory
effectively as a two dimensional non-local, non-relativistic field theory which we do not intend 
to do here. For further discussions on this theory see \cite{Smat}).

Now, let us consider ten dimensional type IIB supergravity on the plane-wave background. In order 
to read off the effective coupling of strings, or the effective Newton constant (of course as 
explained above in the two dimensional sense) one should  work out the ``effective'' 
compactification volume in the plane-wave background. As discussed in \cite{Smat} one should insert
a factor of $({1}/{\sqrt{\mu p^+}})^d$, where $d$ is the number of transverse directions, for 
{\it the} plane-wave  background $d=8$. {}From this, noting that in the string frame (and in 
string units) $G_N^{(10)}=g_s^2$,  the two dimensional effective Newton constant, $G_N^{(2)}$ is 
\cite{G7}
\be
G_N^{(2)}=g_s^2 (\mu p^+)^4 .
\ee
In the plane-wave analysis, however, motivated by the dual gauge theory computations, it has been 
more customary to use $g_2$ instead, where $g_2^2\equiv G_N^{(2)}$ and hence
\be
g_2=g_s (\mu p^+)^2.
\ee
In terms of the dual gauge theory parameters \cite{review},
\be\label{g2}
(\mu p^+)^2 g_s = \frac{J^2}{N} \ ,
\ee
and hence $g_2=J^2/N$. Performing direct dual gauge theory calculations 
it has been confirmed that $g_2$ is indeed the genus counting parameter in the BMN sector 
({\it i.e.} the sector  of ${\cal N}=4$ $U(N)$ SYM gauge theory which consists of operators 
carrying $J$ units of R-charge). 



\begin{thebibliography}{4}

\bibitem{McGreevy:2000cw}
J.~McGreevy, L.~Susskind and N.~Toumbas,
``Invasion of the giant gravitons from Anti-de Sitter space,''
JHEP {\bf 0006}, 008 (2000)
[arXiv:hep-th/0003075].
                                                                                                  
\bibitem{Grisaru:2000zn}
M.~T.~Grisaru, R.~C.~Myers and O.~Tafjord,
``SUSY and Goliath,''
JHEP {\bf 0008}, 040 (2000)
[arXiv:hep-th/0008015].
                                                                                                  
\bibitem{Hashimoto:2000zp}
A.~Hashimoto, S.~Hirano and N.~Itzhaki,
``Large branes in AdS and their field theory dual,''
JHEP {\bf 0008}, 051 (2000)
[arXiv:hep-th/0008016].
                                                                                                  
\bibitem{Das:2000st}
S.~R.~Das, A.~Jevicki and S.~D.~Mathur,
``Vibration modes of giant gravitons,''
Phys.\ Rev.\ D {\bf 63}, 024013 (2001)
[arXiv:hep-th/0009019].

R.~C.~Myers and O.~Tafjord,
``Superstars and giant gravitons,''
JHEP {\bf 0111}, 009 (2001)
[arXiv:hep-th/0109127].
                                                                                                  
F.~Leblond, R.~C.~Myers and D.~C.~Page,
``Superstars and giant gravitons in M-theory,''
JHEP {\bf 0201}, 026 (2002)
[arXiv:hep-th/0111178].

P.~Ouyang,
``Semiclassical quantization of giant gravitons,''
arXiv:hep-th/0212228.

B.~Janssen, Y.~Lozano and D.~Rodriguez-Gomez,
``A microscopical description of giant gravitons. II: The $AdS_5\times S^5$
background,''
Nucl.\ Phys.\ B {\bf 669}, 363 (2003)
[arXiv:hep-th/0303183].


\bibitem{Balasubramanian:2001nh}
V.~Balasubramanian, M.~Berkooz, A.~Naqvi and M.~J.~Strassler,
``Giant gravitons in conformal field theory,''
JHEP {\bf 0204}, 034 (2002)
[arXiv:hep-th/0107119].
                                                                                                  
S.~Corley, A.~Jevicki and S.~Ramgoolam,
``Exact correlators of giant gravitons from dual N = 4 SYM theory,''
Adv.\ Theor.\ Math.\ Phys.\  {\bf 5}, 809 (2002)
[arXiv:hep-th/0111222].
                                                                                                  
D.~Berenstein,
``Shape and holography: Studies of dual operators to giant gravitons,''
Nucl.\ Phys.\ B {\bf 675}, 179 (2003)
[arXiv:hep-th/0306090].

                                                                                                  
\bibitem{Balasubramanian:2002sa}
V.~Balasubramanian, M.-X.~Huang, T.~S.~Levi and A.~Naqvi,
``Open strings from N = 4 super Yang-Mills,''
JHEP {\bf 0208}, 037 (2002)
[arXiv:hep-th/0204196].
                                                                                                  
                                                                                                  
                                                                                                  
                                                                                                  
                                                                                                  
                                                                                                  
                                
\bibitem{juan}
J.~C.~Plefka,
``Lectures on the plane-wave string / gauge theory duality,''
Fortsch.\ Phys.\  {\bf 52}, 264 (2004)
[arXiv:hep-th/0307101].

J.~M.~Maldacena,``TASI 2003 lectures on AdS/CFT,''
arXiv:hep-th/0309246.

M.~Spradlin and A.~Volovich,
``Light-cone string field theory in a plane wave,''
arXiv:hep-th/0310033.




\bibitem{review}
D.~Sadri and M.~M.~Sheikh-Jabbari,
``The plane-wave / super Yang-Mills duality,''
[arXiv:hep-th/0310119].

\bibitem{Hh}
D.~Sadri and M.~M.~Sheikh-Jabbari,
``Giant hedge-hogs: Spikes on giant gravitons,''
Nucl.\ Phys.\ B {\bf 687}, 161 (2004) [arXiv:hep-th/0312155].

\bibitem{MT}
R.~R.~Metsaev and A.~A.~Tseytlin,
``Exactly solvable model of superstring in plane wave Ramond-Ramond background,''
Phys.\ Rev.\ D {\bf 65}, 126004 (2002)
[arXiv:hep-th/0202109].



\bibitem{Bound}
E.~Gava, K.~S.~Narain and M.~H.~Sarmadi,
``On the bound states of p- and (p+2)-branes,''
Nucl.\ Phys.\ B {\bf 504}, 214 (1997)
[arXiv:hep-th/9704006].

M.~M.~Sheikh-Jabbari,
``More on mixed boundary conditions and D-branes bound states,''
Phys.\ Lett.\ B {\bf 425}, 48 (1998)
[arXiv:hep-th/9712199].

\bibitem{SW}
N.~Seiberg and E.~Witten,
``String Theory and Noncommutative Geometry,''
JHEP {\bf 9909}, 032 (1999)
[arXiv:hep-th/9908142].

\bibitem{Lunin}
O.~Lunin, S.~D.~Mathur, I.~Y.~Park and A.~Saxena,
``Tachyon condensation and 'bounce' in the D1-D5 system,''
Nucl.\ Phys.\ B {\bf 679}, 299 (2004)
[arXiv:hep-th/0304007].

J.~M.~Camino and A.~V.~Ramallo,
``Giant gravitons with NSNS B field,''
JHEP {\bf 0109}, 012 (2001)
[arXiv:hep-th/0107142].

\bibitem{Mikhailov}
A.~Mikhailov,
``Giant gravitons from holomorphic surfaces,''
JHEP {\bf 0011}, 027 (2000) [arXiv:hep-th/0010206];
``Nonspherical giant gravitons and matrix theory,'' [arXiv:hep-th/0208077].

D.~Bak, ``Supersymmetric branes in PP wave background,''
Phys.\ Rev.\ D {\bf 67}, 045017 (2003)
[arXiv:hep-th/0204033].




\bibitem{AAS}
F.~Ardalan, H.~Arfaei and M.~M.~Sheikh-Jabbari,
``Noncommutative geometry from strings and branes,''
JHEP {\bf 9902}, 016 (1999)
[arXiv:hep-th/9810072].

C.~S.~Chu and P.~M.~Ho,
``Noncommutative open string and D-brane,''
Nucl.\ Phys.\ B {\bf 550}, 151 (1999)
[arXiv:hep-th/9812219].


\bibitem{ShJ}
M.~R.~Douglas and C.~M.~Hull,
``D-branes and the noncommutative torus,''
JHEP {\bf 9802}, 008 (1998)
[arXiv:hep-th/9711165].

M.~M.~Sheikh-Jabbari,
``Super Yang-Mills theory on noncommutative torus from open strings interactions,''
Phys.\ Lett.\ B {\bf 450}, 119 (1999)
[arXiv:hep-th/9810179].


\bibitem{Lambda-sym}
M.~M.~Sheikh-Jabbari,
``A note on the deformation of Lambda-symmetry in B-field background,''
Phys.\ Lett.\ B {\bf 477}, 325 (2000)
[arXiv:hep-th/9910258].

N.~Seiberg,
``A note on background independence in noncommutative gauge theories,  matrix model and tachyon 
condensation,''
JHEP {\bf 0009}, 003 (2000)
[arXiv:hep-th/0008013].


\bibitem{Ydri}
B.~Ydri, ``Fuzzy physics,''
arXiv:hep-th/0110006.


\bibitem{DSV}
K.~Dasgupta, M.~M.~Sheikh-Jabbari and M.~Van Raamsdonk,
``Protected multiplets of M-theory on a plane wave,''
JHEP {\bf 0209}, 021 (2002) [arXiv:hep-th/0207050].


\bibitem{MV}
J.~McGreevy and H.~Verlinde,
``Strings from tachyons: The c = 1 matrix reloaded,''
JHEP {\bf 0312}, 054 (2003)
[arXiv:hep-th/0304224].




\bibitem{BMN}
D. Berenstein, J. Maldacena, H. Nastase, ``Strings in flat space and pp
waves from ${\cal N}=4$ Super Yang Mills,'' {\it JHEP} {\bf 0204} (2002)
013, hep-th/0202021.


\bibitem{G7}
N.~R.~Constable, D.~Z.~Freedman, M.~Headrick, S.~Minwalla, L.~Motl, A.~Postnikov and W.~Skiba,
``PP-wave string interactions from perturbative Yang-Mills theory,''
JHEP {\bf 0207}, 017 (2002), {\tt arXiv:hep-th/0205089}.

C.~Kristjansen, J.~Plefka, G.~W.~Semenoff and M.~Staudacher,
``A new double-scaling limit of N = 4 super Yang-Mills theory and PP-wave
strings,'' Nucl.\ Phys.\ B {\bf 643}, 3 (2002)
[arXiv:hep-th/0205033].


\bibitem{Smat}

D.~Bak and M.~M.~Sheikh-Jabbari,
``Strong evidence in favor of the existence of S-matrix for strings in plane waves,''
JHEP {\bf 0302}, 019 (2003), {\tt arXiv:hep-th/0211073}.

                                                                                                  


\end{thebibliography}
\end{document}